\UseRawInputEncoding

\documentclass[twocolumn]{aastex62}

\graphicspath{{./}{figures/}}

\submitjournal{ApJL}

\shorttitle{Disk Formation Efficiency \& TDE Rates}
\shortauthors{T.H.T. Wong et al.}

\newcommand{\citeg}[1]{\citep[e.g.,][]{#1}}

\usepackage{amssymb}
\usepackage{amsthm}
\usepackage{amsmath}
\usepackage{natbib}

\begin{document}

\title{Revisiting the Rates and Demographics of Tidal Disruption Events: Effects of the Disk Formation Efficiency}

\author[0000-0001-5570-0926]{Thomas Hong Tsun Wong}\thanks{twht@connect.hku.hk}
\affiliation{Department of Physics, The University of Hong Kong \\
Pokfulam Road, Hong Kong, China}

\author[0000-0003-0841-5182]{Hugo Pfister}
\affiliation{Department of Physics, The University of Hong Kong \\
Pokfulam Road, Hong Kong, China}
\affiliation{DARK, Niels Bohr Institute, University of Copenhagen \\
Jagtvej 128, 2200 København, Denmark}

\author[0000-0002-9589-5235]{Lixin Dai}\thanks{lixindai@hku.hk}
\affiliation{Department of Physics, The University of Hong Kong \\
Pokfulam Road, Hong Kong, China}

\begin{abstract}
Tidal disruption events (TDEs) are valuable probes of the demographics of supermassive black holes as well as the dynamics and population of stars in the centers of galaxies. In this Letter, we focus on studying how the debris disk formation and circularization processes can impact the possibility of observing prompt flares in TDEs.  First, we investigate how the efficiency of disk formation is determined by the key parameters, namely, the black hole mass $M_{\rm BH}$, the stellar mass $m_\star$, and the orbital penetration parameter $\beta$ that quantifies how close the disrupted star would orbit around the black hole.  Then we calculate the intrinsic differential TDE rate as a function of these three parameters.  Combining these two results, we find that the rates of TDEs with prompt disk formation are significantly suppressed around lighter black holes, which provides a plausible explanation for why the observed TDE host black hole mass distribution peaks between $10^6$ and $10^7\,M_\odot$.  Therefore, the consideration of the disk formation efficiency is crucial for recovering the intrinsic black hole demographics from TDEs.  Furthermore, we find that the efficiency of the disk formation process also impacts the distributions of both stellar orbital penetration parameter and stellar mass observed in TDEs.  
\end{abstract}



\section{Introduction \label{sec:intro}}
\noindent Stellar tidal disruption events (TDEs) occur whenever stars are gravitationally scattered into the vicinity of supermassive black holes (SMBHs) where tidal force dominates over the self-gravity of the stars \citep{1988Natur.333..523R}. This happens generally when the star enters the tidal disruption radius of the black hole (BH): \begin{equation}
    r_T \simeq \left(\eta^2\frac{M_{\rm BH}}{m_\star}\right)^{1/3}r_\star  \label{eq:rT}
\end{equation}
where $M_{\rm BH}$ is the BH mass, $m_\star$ and $r_\star$ are the mass and radius of the star.  $\eta$ is a parameter which depends on the details of stellar structure \citep{2020ApJ...905..141L, 2020ApJ...904...98R},  and throughout this work we take $\eta$ to be unity for simplicity.  After the disruption, approximately half of the stellar debris remains bound and forms an accretion disk, while the remaining stellar mass leaves the SMBH.  When a sufficiently large fraction of the star is disrupted and accreted onto the BH, luminous flares can be produced through which we can peek into the event and its host SMBH \citep{2013ApJ...767...25G, 2019ApJ...882L..25L}.  
\begin{figure}
    \includegraphics[width=1.00\linewidth]{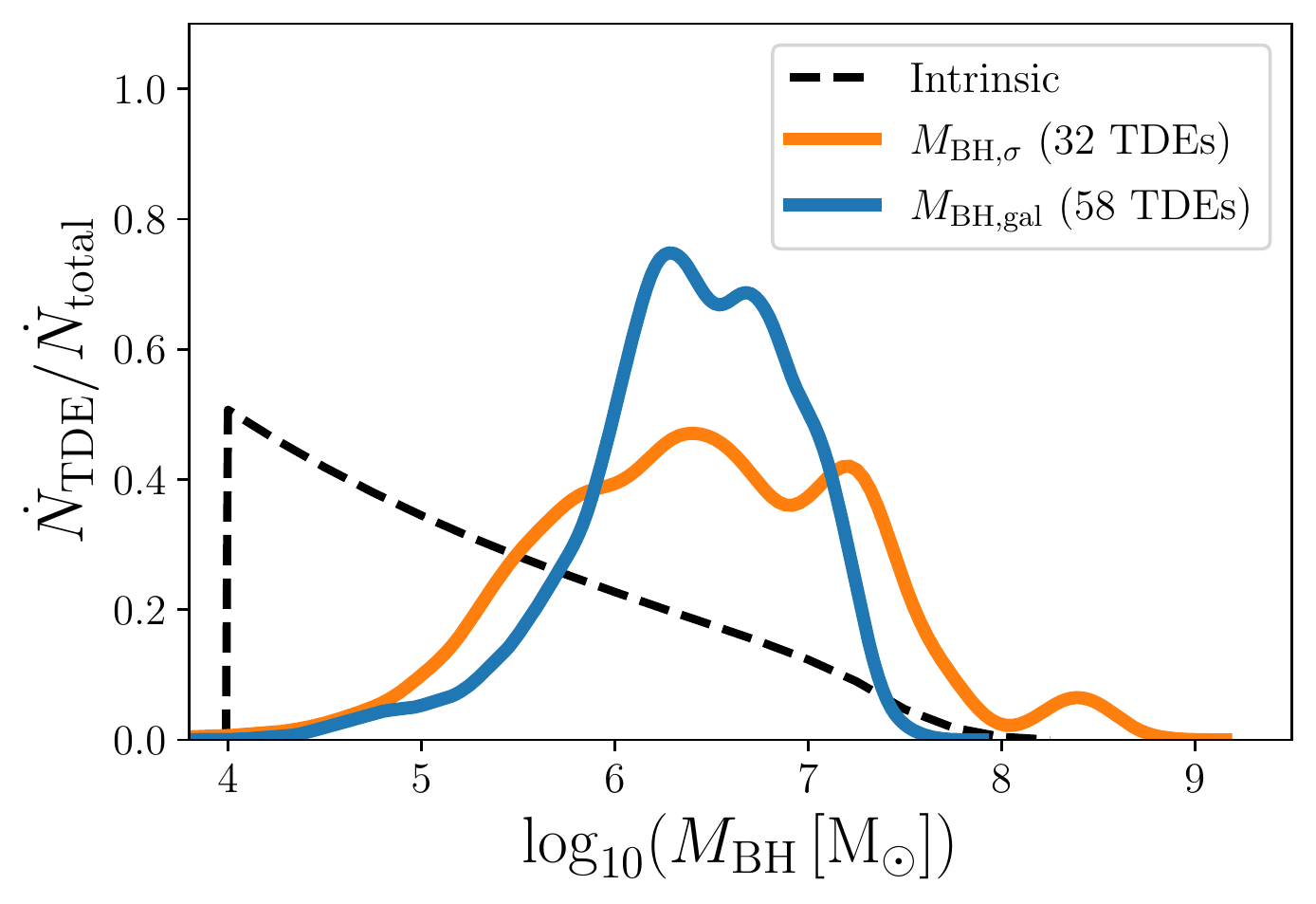}
    \caption{\textbf{The host $M_{\rm BH}$ distribution of the observed TDE candidates compared to the intrinsic distribution from the LC theory.} The solid lines represent our chosen observed TDE candidates, while the colors indicate the methods used to constrain $M_{\rm BH}$ (orange:\ $M_{\rm BH}-\sigma$; blue:\ $M_{\rm BH}-M_{\rm gal}$).   Similar to the bootstrapping method, error bars of individual TDEs are modeled as a Gaussian distribution with the scattering from the $M_{\rm BH}$-scaling relations and the measurement errors from $\sigma$ and $M_{\text{gal}}$.  Both observed TDE $M_{\rm BH}$ distributions center around $10^6-10^7\,M_\odot$, while the intrinsic TDE $M_{\rm BH}$ distribution derived from the LC dynamics (black dashed curve) peaks towards lighter BHs.  The small peak of the orange curve at $10^{8-9}\,M_\odot$ is due to a single event ASASSN-15lh.
    \label{fig:MBHdistGG}}
\end{figure}

The study of TDEs has inspired great interest, as these events can allow us to probe the vast majority of SMBHs which are dormant. Moreover, it is shown that the observables of these events can be used to measure the BH mass, constrain the BH spin, as well as probe the stellar mass and structure \citeg{2012PhRvD..85b4037K, 2019ApJ...872..151M, 2020ApJ...904...73R}. Recently, this field has been growing fast not only because several dozens of TDE candidates have been detected in optical, UV, and X-ray wavebands \citep[see reviews by][]{2020SSRv..216...85S, 2020SSRv..216..124V, 2021arXiv210414580G}, but also because thousands of TDEs are expected to be observed in the next decade using telescopes such as the Vera Rubin Observatory, \emph{eROSITA}, and \emph{Einstein Probe}.  It will therefore be possible to use a large collection of TDEs to construct the demographics of massive BHs.  

The rates of TDEs happening in galaxies with specific stellar density profiles can be calculated using the loss cone (LC) theory, which captures how two-body interactions can bring stars into low angular momentum orbits with pericenter within $r_T$ \citeg{2013CQGra..30x4005M}.  For example, \citet{2004ApJ...600..149W} apply the LC theory upon two common galaxy density profiles and find that the TDE rate should be around $10^{-3} - 10^{-4}\,\rm{gal}^{-1}\,\rm{yr}^{-1}$, which gradually decreases with increasing $M_{\rm BH}$.  \citet{2016MNRAS.455..859S} parameterize the observed galaxy samples with further consideration of stellar mass functions and obtain a similar TDE rate dependency on $M_{\rm BH}$. This trend of decreasing in TDE rate with increasing $M_{\rm BH}$ is seen again while taking into account nuclear star clusters in the center of dwarf galaxies \citep{2020MNRAS.497.2276P, 2021arXiv210305883P}.  The TDE rates also have a close correlation with the merger history \citep{2019MNRAS.488L..29P} and the structures of the host galaxies \citep{2019ApJ...882L..25L, 2020SSRv..216...32F}.  

In summary, while the details depend somewhat on the stellar density profile and galaxy structure, the TDE rate of a galaxy should negatively correlate with $M_{\rm BH}$.  Given there are more smaller galaxies hosting lighter BHs than the opposite \citep{2015ApJ...813...82R}, we expect there should be more TDEs produced around lighter BHs.  However, looking at the observed population of TDEs, one finds very few TDE candidates from BHs with $M_{\rm BH} \lesssim 10^5\,M_\odot$ \citep{2017MNRAS.471.1694W, 2019MNRAS.487.4136W}.  Furthermore, \citet{2016MNRAS.455..859S} and \citet{2017ApJ...842...29H} show that the TDE host $M_{\rm BH}$ distribution peaks between $10^6-10^7\,M_\odot$.  This promotes the proposal that TDEs around more massive BHs have higher chances to be observed.  
Here we re-examine the observed TDE $M_{\rm BH}$ distribution given that the sample size has grown significantly in the past few years.  We refer to several recent reviews and pick out a total of 65 likely TDE candidates (the selection criteria are detailed in Appendix \ref{appsec:TDEdist}).  We also calculate their $M_{\rm BH}$ consistently using two different methods: $M_{\rm BH}-\sigma$ \citep{2001ASPC..249..335M} and $M_{\rm BH}-M_{\rm gal}$ \citep{2015ApJ...813...82R}.  The full list of the selected TDE candidates and their $M_{\rm BH}$ are shown in Table \ref{TDEtable}.  We show the normalized distribution of $M_{\rm BH}$ for these observed TDEs in Fig.\ref{fig:MBHdistGG}.  The distributions for both TDE lists still center around $10^6-10^7\,M_\odot$ and drop off towards both ends of the mass spectrum, regardless of the methods used to derive $M_{\rm BH}$.  In the same figure, we also show the theoretically computed TDE $M_{\rm BH}$ distribution from the LC dynamics (see Appendix \ref{appsec:LC} for detailed calculations) for comparison, where one sees an increasing TDE rate with decreasing $M_{\rm BH}$.

The large discrepancy between the intrinsic and observed TDE $M_{\rm BH}$ distribution can result from many factors such as survey constraints and dusts \citep{2018ApJ...852...72V, 2021ApJ...910...93R}.  Also the disk formation, accretion, and emission processes in TDEs \citep[][respectively]{2021SSRv..217...16B, 2021SSRv..217...12D, 2020SSRv..216..114R} can play a major role in altering the observed TDE $M_{\rm BH}$ distribution.  One possible explanation is that the TDE luminosity is capped at the Eddington luminosity limit and therefore scales with $M_{\rm BH}$ \citep{2016MNRAS.461..371K}.  However, the peak luminosity of the observed TDE candidates do not show any clear linear trend with the host $M_{\rm BH}$ (Fig.\ref{fig:Lbb}). Although we cannot rule out the Eddington-limit model due to possible observational biases and the limited dynamic range of the data, instead, we turn our focus to the disk formation process, since recent hydrodynamics simulations show that not all TDEs can form circular disks efficiently through stream self-crossings \citep[see review by][]{2021SSRv..217...16B}.  In particular, \citet{2015ApJ...804...85S} show that in the case of inefficient disk circularization, the accretion onto SMBHs will happen on timescales much longer than the fallback timescale, which can lower the possibility of the detection of such events.  Based on these simulations, \citet{2015ApJ...812L..39D} and \citet{2015ApJ...809..166G} use first-order calculations to quantify how the disk formation promptness depends on the BH mass and stellar orbital parameters. It is found that in TDEs around more massive BHs the stream self-crossings are closer to the BH due to stronger general relativistic (GR) effects. The closer self-crossings give stronger collisions, leading to faster disk formation and accretion and more prompt flares.  Therefore, the disk formation process is a natural candidate for a mechanism which can promote TDEs around more massive SMBHs to be observed.

In this paper, we aim at mitigating the TDE $M_{\rm BH}$ distribution tension from the perspective of disk formation efficiency.  We hypothesize that inefficient disk circularization processes significantly reduce the chance of TDEs being detected around lighter SMBHs.  Using the first-order framework of \citet{2015ApJ...812L..39D}, we further quantify the disk formation efficiency as a function of key parameters (Section \ref{sec:ssinteract}). Next we obtain the differential TDE rates from the LC theory, and check how including the correction by the disk formation efficiency can shift the TDE $M_{\rm BH}$ distribution (Section \ref{sec:tderate}).  Then we reexamine the distributions of some key parameters in TDEs (Section \ref{sec:dist}). Lastly, a summary and further discussions are given (Section \ref{sec:discussion}).

\section{Efficiency of disk formation from debris stream self-crossing \label{sec:ssinteract}}

\begin{figure*}
\centering
\includegraphics[width=0.95\linewidth]{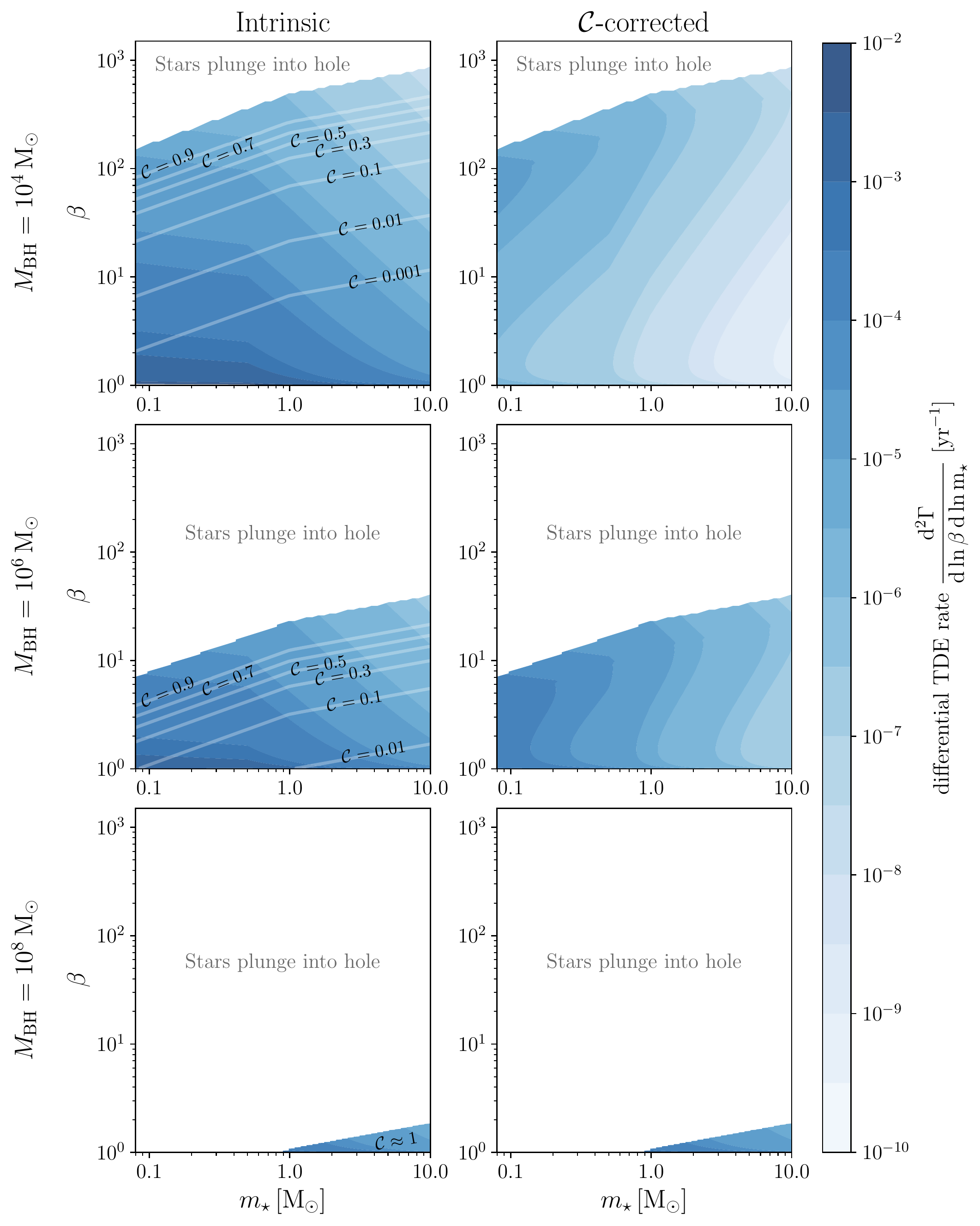}
\caption{\textbf{Comparison between the intrinsic and $\mathcal{C}$-corrected differential TDE rates.}  For $M_{\rm BH}/M_\odot=\{10^4, 10^6, 10^8\}$ (top, middle, bottom rows), the intrinsic and corrected differential TDE rates are shown in the left and right panels, respectively.  The white regions represent the parameter space where stars directly plunge into the BH event horizon.  Contours of disk formation efficiencies $\mathcal{C}$ are overplotted in the left panels.  $\mathcal{C}$ has a monotonically increasing trend towards the upper-left corner, (higher $\beta$ and lower $m_\star$).  One can see that the $\mathcal{C}$-correction leads to a significant drop in TDE rates at low $\beta$ for lighter BHs.  For $M_{\rm BH} = 10^8\,M_\odot$, $\mathcal{C}\approx 1$  across the TDE-allowed parameter space.  \label{fig:circrate}}
\end{figure*}

Recent studies have consistently shown that the formation of TDE disks is mainly caused by the self-crossing of the debris stream due to GR apsidal precession near SMBHs. Some fraction of the stellar debris becomes unbound due to shocks, while the remaining rapidly dissipates its orbital energy to form a somewhat circular accretion disk, generally within a few orbital timescales \citep{2020MNRAS.492..686L}.  For stars disrupted along inclined orbits around spinning BHs, Lense-Thirring precession can complicate the process \citep{1994ApJ...422..508K, 2013ApJ...775L...9D, 2015ApJ...809..166G}, although the latest hydrodynamical simulations show that disks still form after a short delay \citep{2019arXiv191010154L}.  Albeit these advancements, full-scale simulations of the complex disk formation processes are still not achieved due to computational limitations.  

In this study, we follow the first-order calculations in \citet{2015ApJ...812L..39D} for debris dynamics and kinematics around a Schwarzschild BH. The basics are detailed in Appendix \ref{appsec:ssinteract}. Starting from there, the specific energy loss at the self-crossing point $R_I$ is approximated as 
\begin{equation}
    \Delta E_{\text{first}} = \left|\frac{1}{2}v_f^2 - \frac{1}{2}v_i^2\right|, \label{eq:deltaEfirst}
\end{equation}
where $v_f$ and $v_i$ are the debris speeds before and after collision respectively.  The bound debris has to lose a significant amount of orbital energy, likely through the shocks and dissipation happening in repeated self-crossings, in order to form a disk.  In theory, if the debris could completely circularize, the classical circularization radius can be calculated using the conservation of angular momentum: 
\begin{equation}
    r_{\text{circ}} = \frac{2r_T}{\beta} \label{eq:rcirc}
\end{equation}
where $\beta=r_T/r_p$ is the penetration parameter quantifying how deeply the star plunges into the BH gravitational potential well. Therefore, the total specific orbital energy needed to be removed from the initially elliptical orbit for complete circularization is: 
\begin{eqnarray}
    \Delta E_{\text{total}} \,&=&\, E_{\text{ellip}} - E_{\text{circ}} \nonumber \\
    &=&\, -\frac{GM_{\rm BH}}{2a_{\text{mb}}} + \frac{GM_{\rm BH}}{2r_{\text{circ}}} \nonumber \\
    &=&\, \frac{GM_{\rm BH}}{2}\left(\frac{\beta}{2r_T} - \frac{1}{a_{\text{mb}}}\right) \label{eq:deltaEtotal}
\end{eqnarray}
where $a_{\rm mb}$ is the semi-major axis of the most bound debris orbit.  (Strictly speaking, the debris falling back at the peak of the flare should be slightly less bound than the most bound debris. Therefore, the $\Delta E_{\text{total}}$ obtained here is a lower limit.)

Given that the subsequent collisions and interactions after the first debris self-crossing are hard to trace analytically, we define a first-order, dimensionless disk formation (circularization) efficiency parameter by scaling Eq.\ref{eq:deltaEfirst} with Eq.\ref{eq:deltaEtotal}:
\begin{equation}
    \mathcal{C}\left(M_{\rm BH},\beta,m_\star\right) \equiv \frac{\Delta E_{\text{first}}}{\Delta E_{\text{total}}}  \label{eq:circeff}
\end{equation}
When $\mathcal{C}\ll 1$, the debris stream self-crossing happens far away from the BH and the disk formation is expected to be slow, which lowers the possibility of detecting prompt flares from such events.  On the other hand, when $\mathcal{C} \sim 1$, the debris can already dissipate a large fraction of its orbital energy through the shocks at the first self-crossing, and we expect such TDEs will likely produce prompt, observable flares.  

The tidal radius $r_T$ can be expressed as a function of only $M_{\rm BH}$ and $m_\star$ for main-sequence (MS) stars, since their masses and radii are linked by \citet{1990sse..book.....K}: 
\begin{equation}
    \frac{r_\star}{\rm{R}_\odot} = 
    \begin{cases}
        \left(\displaystyle\frac{m_\star}{M_\odot}\right)^{0.8}&\text{for}\,\, m_\star\leq M_\odot \\
        \left(\displaystyle\frac{m_\star}{M_\odot}\right)^{0.57}&\text{for}\,\, m_\star\geq M_\odot
    \end{cases} \label{eq:mrrelation}
\end{equation}
Therefore, $\mathcal{C}$ can be expressed as a function of three parameters: $M_{\rm BH}$, $m_\star$ and $\beta$.  We plot the values of $\mathcal{C}$ as contour lines in the left column of Fig. \ref{fig:circrate} for three representative $M_{\rm BH}$.  One can see that for intermediate-mass black holes (IMBHs) with $M_{\rm BH} \sim 10^4\,M_\odot$, extremely large $\beta >$ few $\times$ 10 will be needed for having efficient disk formation.  For $M_{\rm BH} \sim 10^6\,M_\odot$, low-mass, denser stars with $m_\star \sim 0.1\,M_\odot$ can have moderately prompt disk formation for all $\beta$, while for more massive stars the disk formation is slow unless $\beta >$ few. For very massive black holes with $M_{\rm BH} \sim 10^8\,M_\odot$, the disk formation is always efficient, but only stars with $m_\star \gtrsim 1\,M_\odot$ can still be disrupted outside the BH event horizon.  

In the subsequent sections, we apply the  $\mathcal{C}$-correction to the observed TDE rates and demographics in two different ways:\ one way is to directly use $\mathcal{C}$ as the probability of observing an event $P_{\rm obs}$, the other way is that we pick a particular threshold value $\mathcal{C}_{\text{thres}}$ to screen out TDEs with $\mathcal{C}<\mathcal{C}_{\text{thres}}$.  For the latter, reasonable choices of $\mathcal{C}_{\text{thres}}$ can be obtained by comparing the TDE light curve decay timescale ($\sim1$ year for typical parameters) and the orbital timescale of the most bound orbit ($T_{\rm mb}\sim1$ month for typical parameters). The timescale to circularize the elliptical orbit is approximately $T_{\rm circ} = n T_{\rm mb}$, where $n$ is the number of stream self-crossings \citep{1998AIPC..431..141U}.  For the dynamical fallback pattern to be reserved, the disk formation should in principle happen faster than the fallback timescale, which gives $n\lesssim 10$ and $\mathcal{C}_{\rm thres}\sim 1/n \gtrsim 0.1$.  

We also note that there are ongoing discussions on whether the optical emissions from TDEs are powered by accretion or the shocks induced in the debris self-crossing \citeg{2015ApJ...806..164P,  2020MNRAS.495.1374B}.  Nonetheless, the chance of observing a particular TDE flare, even if only powered by debris self-crossing, should still have a tight correlation with $\Delta E_{\rm first}$.  Therefore, the correction between the observed TDE rates and  $\mathcal{C}$  should hold regardless of the origin of TDE optical emissions, which awaits to be disclosed by further studies.

\section{Revised TDE Rates \label{sec:tderate}}

\subsection{Intrinsic TDE rates from loss cone dynamics \label{subsec:LCrate}}

We apply the LC theory and calculate the TDE rates.  In this work, we only consider bright TDE flares from full disruption scenarios, since it is hard to distinguish whether a dim flare is produced due to slow disk formation or partial disruption.  We refer the readers to \citet{2021arXiv210305883P} for the calculations used in this work, which we also include in Appendix \ref{appsec:LC} for completeness.  

Assuming the Kroupa stellar mass function \citep{2001MNRAS.322..231K}, an isothermal stellar density profile  ($\rho\propto r^{-2}$), and the BH mass following the $M_{\rm BH}-\sigma$ relation \citep{2001ASPC..249..335M}, the intrinsic differential TDE rate can be analytically computed as a function of $M_{\rm BH}$, $m_\star$, and $\beta$ \citep{2021arXiv210305883P}: 
\begin{align}
    \frac{\rm{d}^2\Gamma}{\rm{d}\ln\beta\,\rm{d}\ln m_\star}\,\, =&\,\, \frac{8\pi^2GM_{\rm BH} r_T}{\beta}\phi(m_\star)m_\star\times \label{eq:diffrate1} \\
    &\int^{GM_{\rm BH}/r_T}_{0}\mathcal{G}(E,M_{\rm BH},\beta,m_\star)\rm{d}E \nonumber
\end{align}
where $\phi(m_\star)$ is the Kroupa stellar mass function and $\mathcal{G}(E,M_{\rm BH},\beta,m_\star)$ is a complex function that gathers the essence of the LC dynamics.  

Under this simplified framework, we vary $M_{\rm BH}$ in the range $\left[10^4\,M_\odot, 10^9\,M_\odot\right]$ while bearing in mind the existence of IMBHs is still under debate. We include the MS stars in range of $\left[0.08\,M_\odot,10\,M_\odot\right]$. The penetration parameter $\beta$ varies between $1$ and $\beta_{\rm max}$, where $\beta_{\text{max}}=r_T/2r_g$ corresponds to the orbit with $r_p$ at the BH Schwarzschild radius.  For a fixed $M_{\rm BH}$, $\beta_{\text{max}}$ is set by the most massive star, so we have $\beta_{\rm max}^{10\,M_\odot}=\beta_{\rm max}(M_{\rm BH},10\,M_\odot)$, which has a value of $\sim10^3$ when $M_{\rm BH}=10^4\,M_\odot$, $\sim 40$ when $M_{\rm BH}=10^6\,M_\odot$, and $\sim 1$ when $M_{\rm BH}=10^8\,M_\odot$. We show the intrinsic differential TDE rate in Fig.\ref{fig:circrate} (left panels). This plot depicts that most TDEs have $\beta\sim1$, and most disrupted stars have low masses unless for $M_{\rm BH}\gtrsim 10^8\,M_\odot$.

\subsection{Corrected TDE rates with disk formation efficiency \label{subsec:circrate}}
Following the discussion in Sec.\ref{sec:ssinteract} we multiply  the intrinsic differential TDE rates (Eq.\ref{eq:diffrate1}) with the disk formation efficiency parameter $\mathcal{C}$ to obtain the predicted observed differential TDE rates: 
\begin{equation}
    \frac{\rm{d}^2\Gamma_{\mathcal{C}}}{\rm{d}\ln \beta\,\rm{d}\ln m_\star}\, \equiv\,\, \mathcal{C}\times\frac{\rm{d}^2\Gamma}{\rm{d}\ln \beta\,\rm{d}\ln m_\star} \label{eq:circrate}
\end{equation}
where $\Gamma$ and $\Gamma_{\mathcal{C}}$ stand for the intrinsic and $\mathcal{C}$-corrected rates respectively. $\Gamma_{\mathcal{C}}$, as a function of $\beta$ and $m_\star$, is shown in Fig.\ref{fig:circrate} (right panels). 
Comparing the two rates, the most notable difference is that the $\beta$-distribution becomes much more even for the same $m_\star$ after the $\mathcal{C}$ correction.  This is because the substantial reduction of the TDE rates mostly happens at low $\beta$ due to small $\mathcal{C}$ where the intrinsic rate is large. Also, $\Gamma_{\mathcal{C}}$ has a more sensitive dependence on $m_\star$, which results from that larger stars which are rare initially also generally have smaller $\mathcal{C}$ values.  However, for $M_{\rm BH} \approx 10^8\,M_\odot$, $\mathcal{C}\approx 1$, so the rates and distributions are barely altered.

\subsection{TDE host black hole mass distribution \label{subsec:totalrate}}

\begin{figure*}
\centering
\includegraphics[width=1.0\linewidth]{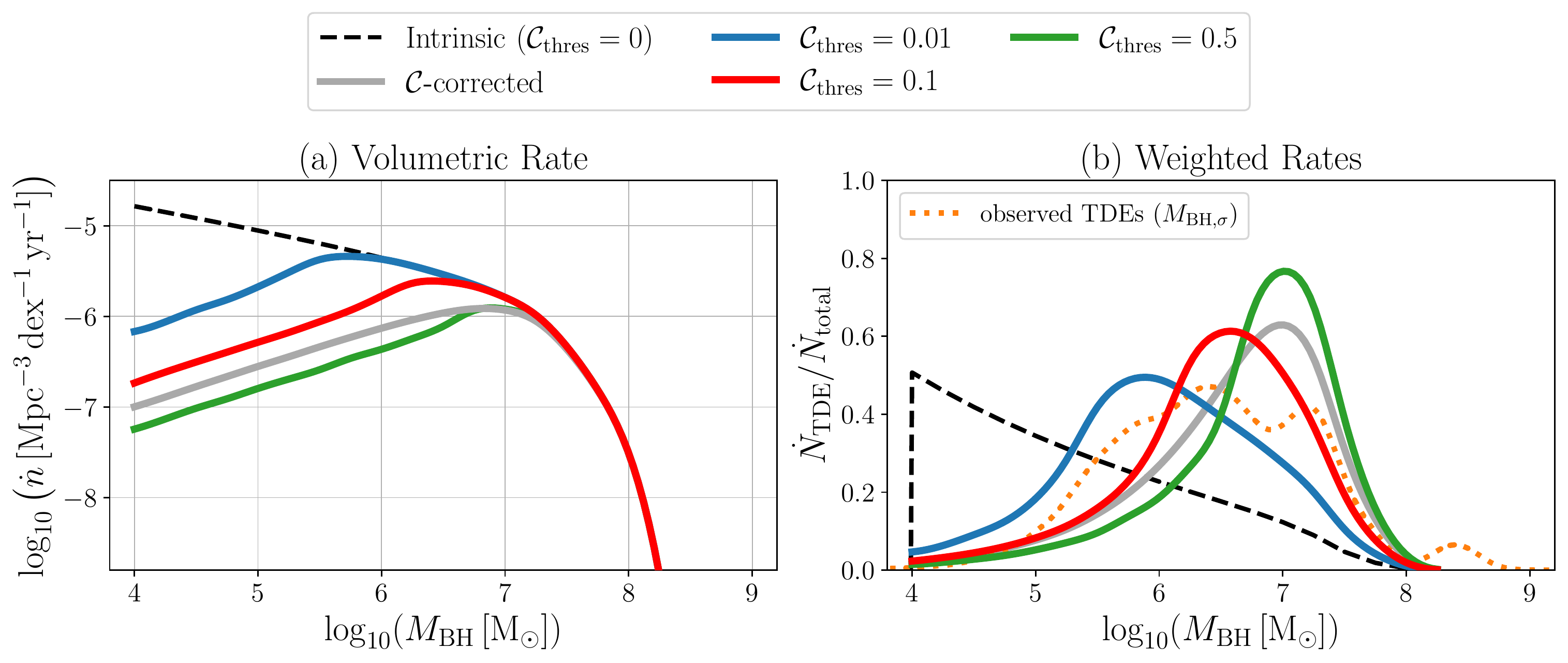}
\caption{\textbf{The intrinsic and $\mathcal{C}$-corrected volumetric TDE rates and $M_{\rm BH}$ distributions.} Panel (a) shows the TDE volumetric rates as a function of $M_{\rm BH}$, with different ways of implementing the disk formation efficiency correction.  Panel (b) shows how the value of $\mathcal{C}_{\rm thres}$ affects the TDE $M_{\rm BH}$ distribution. The intrinsic and observed distributions (orange dotted: $M_{\rm BH}-\sigma$) same as in Fig.\ref{fig:MBHdistGG} are included for comparison. The distribution with $\mathcal{C}_{\text{thres}}=0.1$ (red solid) best replicates the median of the observed TDE samples.  Overall, the rate suppression around light SMBHs due to slow disk formation is evident in both plots.  \label{fig:totalrate}}
\end{figure*}

The volumetric TDE rates as a function of the host BH mass can be obtained with the results above and the information of the black hole mass function (BHMF).  The BHMF $\Phi_{\rm BH}$, defined as the number of SMBHs per co-moving volume between masses $M_{\rm BH}$ and $M_{\rm BH}+dM_{\rm BH}$, provides knowledge on the growth and evolution of SMBHs and galaxies.  Currently, the observationally constrained BHMF still has a large uncertainty \citeg{2016MNRAS.460.3119S, 2019ApJ...883L..18G}.  In particular, the uncertainty in BH occupation fraction increases significantly towards lighter BHs of $M_{\rm BH} \lesssim 10^5\,M_\odot$, which also results in uncertainty of TDE rates around such BHs.  We have shown in Appendix \ref{appsec:BHMF} that the uncertainty in the moderately constrained BHMF at the low-mass end is insufficient to explain the TDE BH mass distribution as observed.

In this work, we adopt the BHMF $\Phi_{\rm BH}$ derived by \citet{2019ApJ...883L..18G}\footnote{The equation is the corrected version of \citet{2019ApJ...883L..18G}, reported in \citet{2021arXiv210305883P}.}:
\begin{equation}
\begin{split}
    \log_{10}\left(\frac{\Phi_{\rm BH}}{\text{Mpc}^{-3}\,M_\odot^{-1}}\right) = &-9.82 - 1.10\times\log_{10}\left(\frac{M_{\rm BH}}{10^7\,M_\odot}\right)\\
    &- \left(\frac{M_{\rm BH}}{128\times10^7\,M_\odot}\right)^{1/\ln(10)} \label{eq:GalloBHMF}
\end{split}
\end{equation}
which constrains the local SMBH occupation fraction and incorporates the latest galaxy stellar mass function with \emph{Chandra} X-ray data.  Most importantly, this BHMF takes into account the non-unity SMBH occupation fraction in galaxies towards the near-IMBH mass spectrum.  Furthermore, since all TDE candidates discovered so far are at low redshifts $z \sim 0.01 - 0.4$, this particular BHMF derived from the local SMBHs is applicable for this redshift range.  

The predicted volumetric intrinsic and $\mathcal{C}$-corrected TDE rates are respectively: 
\begin{subequations}
\begin{align}
    \dot{n}(M_{\rm BH}) &= \Phi_{{\rm BH}}\times\int \left(\frac{\rm{d}^2\Gamma}{\rm{d}\ln\beta\,\rm{d}\ln m_\star}\right) \rm{d}\ln\beta\,\rm{d}\ln m_\star  \label{eq:totalrate} \\
    \dot{n}_{\mathcal{C}}(M_{\rm BH}) &= \Phi_{{\rm BH}}\times\int \left(\frac{\rm{d}^2\Gamma_{\mathcal{C}}}{\rm{d}\ln\beta\,\rm{d}\ln m_\star}\right) \rm{d}\ln\beta\,\rm{d}\ln m_\star \label{eq:totalcircrate}
\end{align}
\end{subequations}
which are plotted in Fig.\ref{fig:totalrate}(a). The integrated volumetric rate of all $M_{\rm BH}$ is $\lesssim 10^{-5}\,\rm{Mpc}^{-3}\,\rm{yr}^{-1}$ for $\mathcal{C}_{\rm thres}\gtrsim 0.1$, which lies a little beyond the TDE rate inferred from current observations $\dot{n}\approx 4\times10^{-8.4} - 5.4\times10^{-6}\,\rm{Mpc}^{-3}\,\rm{yr}^{-1}$ (\citealt{2020SSRv..216...35S} and references therein). This could result from the isothermal stellar density profile used in our calculation, the BHMF uncertainty at low $M_{\rm BH}$, and observational limitations.  Therefore, we check how the normalized TDE $M_{\rm BH}$ distribution changes after considering the disk formation efficiency in Fig.\ref{fig:totalrate}(b), which depends less on various assumptions.    Similar to the rates, after applying the $\mathcal{C}$ corrections, the $M_{\rm BH}$ distribution shifts towards the more massive side of the spectrum and peaks around $10^{6-7}\,M_\odot$. 

Suppose that the disk formation efficiency is the sole mechanism affecting the observed $M_{\rm BH}$ distribution, we find $\mathcal{C}_{\rm thres} \approx 0.1$ would give the best-fit distribution according to the centroid values at $M_{\rm BH}\approx10^{6.5}\,M_\odot$. 
This suggests that current transient surveys favor the detection of TDEs with moderately prompt disk formation.

\section{Distributions of TDE: Penetration Parameter and Stellar Mass \label{sec:dist}}

\subsection{Distribution of the penetration parameter $\beta$ \label{subsec:betadist}}
\begin{figure}
    \centering
    \includegraphics[width=1.0\linewidth]{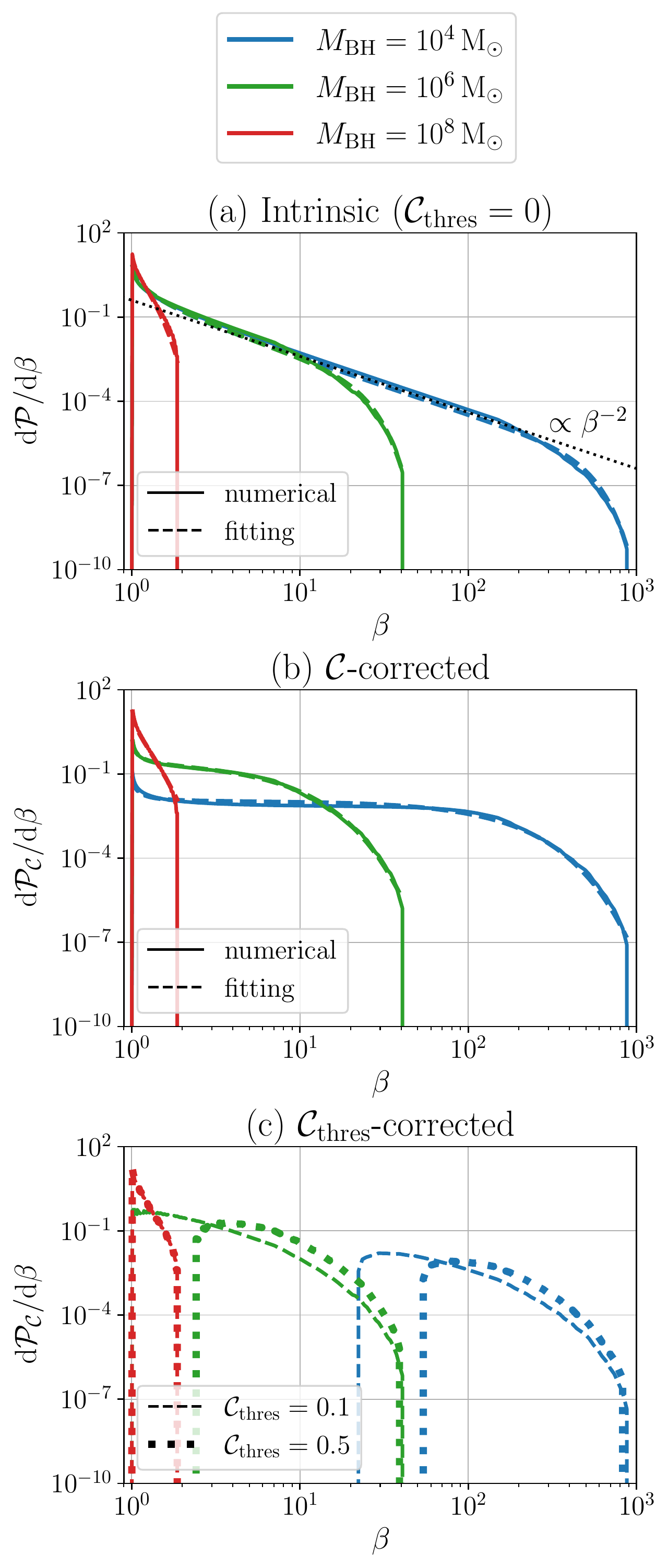}
    \caption{\textbf{Numerical and empirically-fitted $\beta$-distributions derived from intrinsic and  $\mathcal{C}$-corrected TDE rates}.  Different colors indicate different $M_{\rm BH}=10^4\,M_\odot\,(\text{blue}),\,10^6\,M_\odot\,(\text{green}),\,10^8\,M_\odot\,(\text{red})$. In panels (a) and (b), empirical fitting formulae of Eq.\ref{eq:betadist_fit} and \ref{eq:circbetadist_fit} are plotted in dashed lines with their corresponding numerical results in solid lines.
    In panel (c), two different $\mathcal{C}_{\text{thres}}=0.1\,(\text{dashed})$ and $0.5\,(\text{dotted})$ are included.}
    \label{fig:betadist}
\end{figure}

The intrinsic $\mathcal{P}(\beta)$ and $\mathcal{C}$-corrected $\mathcal{P}_{\mathcal{C}}(\beta)$ $\beta$ probability density functions can be numerically computed respectively as:
\begin{subequations}
\begin{align}
    \frac{\rm{d}\mathcal{P}}{\rm{d}\beta} &= \frac{1}{\beta\Gamma}\int\left(\frac{\rm{d}^2\Gamma}{\rm{d}\ln\beta\,\rm{d}\ln m_\star}\right)\rm{d}\ln m_\star \,  \label{eq:betadist} \\
    \frac{\rm{d}\mathcal{P}_{\mathcal{C}}}{\rm{d}\beta} &= \frac{1}{\beta\Gamma_{\mathcal{C}}}\int\left(\frac{\rm{d}^2\Gamma_{\mathcal{C}}}{\rm{d}\ln\beta\,\rm{d}\ln m_\star}\right)\rm{d}\ln m_\star \, \label{eq:circbetadist}
\end{align}
\end{subequations}

We show the intrinsic $\beta$-distributions (Fig.\ref{fig:betadist}(a)) for three representative $M_{\rm BH}$.  For $1 \ll \beta \ll \beta_{\rm max}^{10\,M_\odot}$, we find $d\mathcal{P}/d\beta\propto\beta^{-2}$,  which is consistent with the rate dominated by the full LC regime (see \citealt{2016MNRAS.455..859S}).  After $\beta$ reaches around $\beta_{\rm max}^{0.08\,M_\odot}$ (maximal possible $\beta$ for TDEs of $m_\star=0.08\,M_\odot$), low-mass stars start to plunge into the horizon, resulting in a drop in the slope.  The $\beta$-distributions is cut off at $\beta_{\rm max}^{10\,M_\odot}$ when all stars undergo direct plunging.  On the other side, as $\beta\rightarrow 1$, $d\mathcal{P}/d\beta$ is expected to peak owing to the additional contribution from TDEs in the diffusive regime.  

The $\beta$-distributions with $\mathcal{C}$ correction implemented are shown in Fig.\ref{fig:betadist}(b) and Fig.\ref{fig:betadist}(c).  As expected, low-$\beta$ TDEs are removed, and larger $\mathcal{C}_{\text{thres}}$ values generally yield higher low-$\beta$ cutoffs. Adopting the nominal value of $\mathcal{C}_{\rm thres}=0.1$, we can see that for $M_{\rm BH} \lesssim10^6\,M_\odot$, low-$\beta$ TDEs are significantly reduced.  Furthermore, for IMBHs with $M_{\rm BH} \lesssim10^4\,M_\odot\,(10^5\,M_\odot)$, only TDEs with $\beta\gtrsim20\,(5)$ can have prompt disk formation.  

The numerically obtained $\beta$-distributions in Fig.\ref{fig:betadist}(a,b) could be accurately fitted in log-log space with the following empirical formulae:
\begin{subequations}
    \begin{align}
        \frac{\rm{d}\mathcal{P}}{\rm{d}\beta}\bigg|_{\text{fit}} \propto\,\, &\left[\left(\frac{\beta}{\beta-1}\right)^{0.1} + \left(\frac{\beta}{\beta-1}\right)^{-0.1} - 2\right]\times \nonumber\\
        &\exp\left[-2\left(\frac{\beta}{\beta_{\text{max}}^{10\,M_\odot}}\right)^{2.7}\right] \,\,\text{for}\,\,1\lesssim\beta<\beta_{\text{max}}^{10\,M_\odot}, \label{eq:betadist_fit} \\
        \frac{\rm{d}\mathcal{P}_{\mathcal{C}}}{\rm{d}\beta}\bigg|_{\text{fit}} \propto\,\, &\left[\left(\frac{\beta}{\beta-1}\right)^{0.6} + \left(\frac{\beta}{\beta-1}\right)^{-0.6}\right]\times \nonumber\\
        &\exp\left[-9\left(\frac{\beta}{\beta_{\text{max}}^{10\,M_\odot}}\right)^{1.1}\right] \,\,\text{for}\,\,1\lesssim\beta<\beta_{\text{max}}^{10\,M_\odot},  \label{eq:circbetadist_fit}
    \end{align}
\end{subequations}
\begin{figure*}
\includegraphics[width=1.00\linewidth]{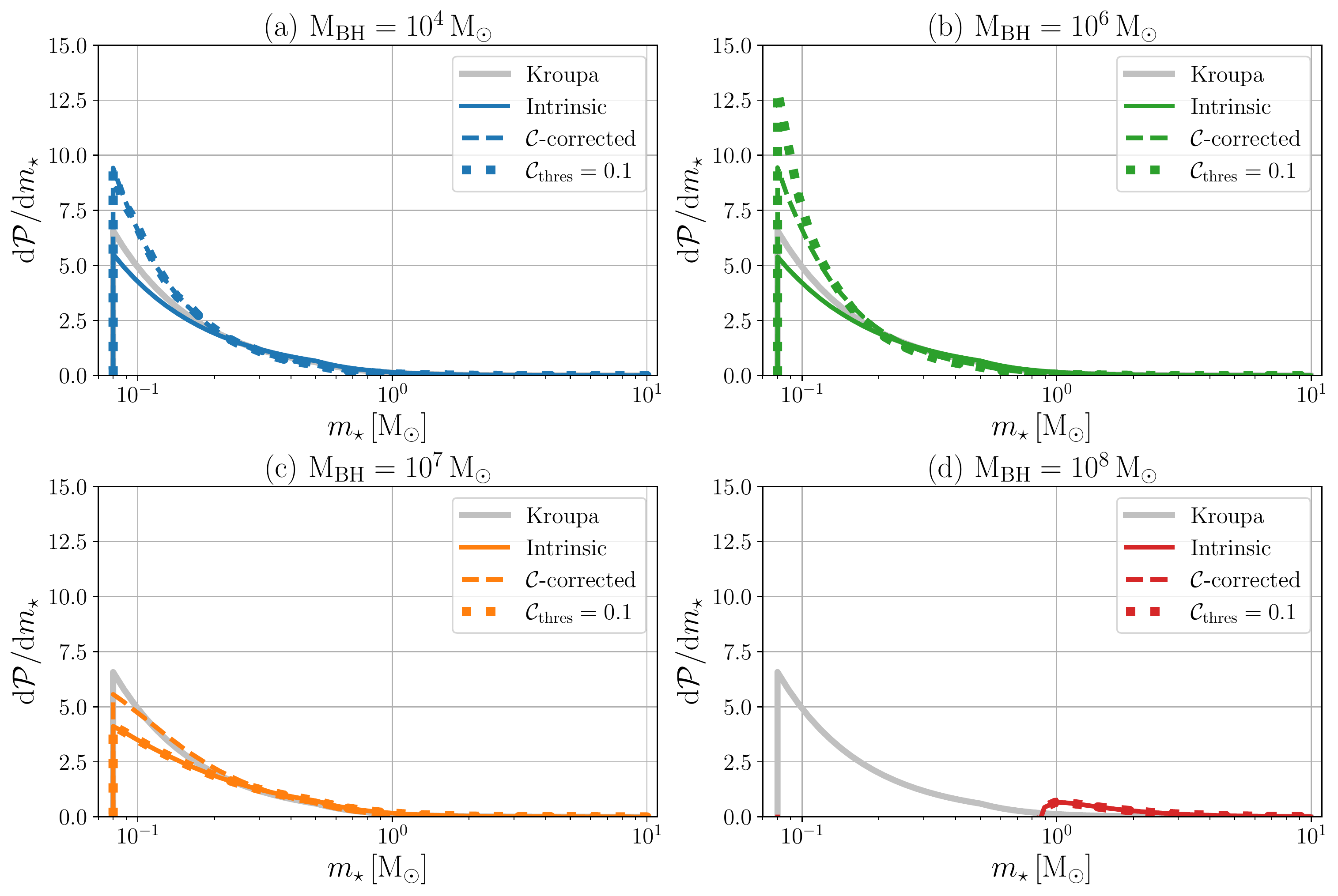}
\caption{\textbf{The stellar mass probability density functions in TDEs before and after the correction with $\mathcal{C}$}.  For four distinct $M_{\rm BH}=10^4\,M_\odot\,(\text{blue}),\,10^6\,M_\odot\,(\text{green}),\,10^7\,M_\odot\,(\text{orange}),\,10^8\,M_\odot\,(\text{red})$, we show the intrinsic (solid), $\mathcal{C}$-corrected (dashed), and $\mathcal{C}_{\rm thres}=0.1$ case (dotted) stellar mass distributions in TDEs. The Kroupa initial mass function  is plotted for comparison (grey solid). Low-mass stars have a lower intrinsic presentation in TDEs due to a combination of the LC physics as well as their direct plunge when $M_{\rm BH}$ is large.  For TDEs with $M_{\rm BH}\gtrsim10^{8}\,M_\odot$, there exists a cutoff of low-mass stars. For TDEs with $M_{\rm BH}\lesssim10^{6}\,M_\odot$, the $\mathcal{C}$-correction can over-compensate the suppression of low-mass stars. 
\label{fig:mstardist}}
\end{figure*}
\noindent which are overplotted on the numerical results. The generic form of the formulae captures the entire behaviour across the full $\beta$ range:\ the rising towards $\beta\approx1$, the asymptotic $\beta^{-2}$ proportionality ($\beta$-independent) in the $1\ll \beta \ll \beta_{\rm max}^{10\,M_\odot}$ regime, and the fast exponential decay while approaching $\beta_{\rm max}^{10\,M_\odot}$, with a maximum error of $0.39$ dex.

\subsection{Distribution of the disrupted stellar mass $m_\star$ \label{subsec:mstardist}}

In a similar fashion, we can calculate the intrinsic $\mathcal{P}(m_\star)$ and $\mathcal{C}$-corrected $\mathcal{P}_{\mathcal{C}}(m_\star)$  stellar probability density functions as: 
\begin{subequations}
\begin{align}
    \frac{\rm{d}\mathcal{P}}{\rm{d}m_\star} &= \frac{1}{m_\star\Gamma}\int\left(\frac{\rm{d}^2\Gamma}{\rm{d}\ln\beta\,\rm{d}\ln m_\star}\right)\rm{d}\ln\beta \, , \label{eq:mstardist} \\
    \frac{\rm{d}\mathcal{P}_{\mathcal{C}}}{\rm{d}m_\star} &= \frac{1}{m_\star\Gamma_{\mathcal{C}}}\int\left(\frac{\rm{d}^2\Gamma_{\mathcal{C}}}{\rm{d}\ln\beta\,\rm{d}\ln m_\star}\right)\rm{d}\ln\beta \, .  \label{eq:circmstardist}
\end{align}
\end{subequations}
Fig.\ref{fig:mstardist} shows the predicted stellar mass distribution for a range of black hole masses. One can see that the intrinsic LC process somewhat suppresses the representation of low-mass stars in TDEs (see \citealt{2021arXiv210305883P} for the explanation).  The suppression is not strong unless when $M_{\rm BH} > 10^7\,M_\odot$ and low-mass stars on high-$\beta$ orbits start to directly plunge. When $M_{\rm BH} \gtrsim10^8\,M_\odot$, one can see all low-mass stars are removed.

The disk formation efficiency, however, is higher for low-mass stars and therefore leads to an enhancement of their presence in the observed TDE population. When putting all the factors together, for the majority TDE population ($M_{\rm BH}\lesssim 10^6 M_\odot$), we should still expect to observe more TDEs from low-mass stars than the initial stellar mass distribution.

\section{Discussion and Summary \label{sec:discussion}}

\noindent In this paper, we consider how the physics of TDE disk formation can affect their observed rates and demographics.  We summarize our findings as below: 

\begin{itemize}
\item We have systematically recalculated the host BH masses for an updated list of observed TDEs using two methods, $M_{\rm BH}-\sigma$ and $M_{\rm BH}-M_{\rm gal}$, based on the observed values of $\sigma$ and $M_{\rm gal}$ from previous literature.  We find that the observed TDE $M_{\rm BH}$ distribution peaks between $10^6-10^7\,M_\odot$, which shows a drastic discrepancy when compared to the intrinsic distribution of TDE $M_{\rm BH}$ obtained using 
the LC dynamics and the BHMF (Fig.\ref{fig:MBHdistGG}).  
\item We address this discrepancy by considering the efficiency of TDE disk formation ($\mathcal{C}$), which is quantified to the first order as the ratio of the energy loss in the first stream self-crossing over the total energy needed to be reduced for complete circularization (Eq.\ref{eq:circeff}).  We find that for BHs with $M_{\rm BH}\lesssim10^6\,M_\odot$, $\mathcal{C}$ is typically very low ($\lesssim 0.1$) for typical TDEs with low $\beta$ (Fig.\ref{fig:circrate}).  
\item While applying $\mathcal{C}$ to correct the TDE rates, we naturally obtain a significant suppression of TDE rates around lighter BHs.   The $M_{\rm BH}$ distribution peak can be shifted to fit the observed one when we only keep the TDEs with $\mathcal{C} \gtrsim0.1$ (Fig.\ref{fig:totalrate}).

\item There are a few interesting consequences induced by including the $\mathcal{C}$ correction.  First, high-$\beta$ TDEs are favored since they have more prompt disk formation.  Therefore, the $\beta$-distribution of the TDEs with prompt disk formation is much flatter than the intrinsic one (Fig.\ref{fig:betadist}). We provide the fitted formulae for both the intrinsic and corrected $\beta$-distributions (Eq.\ref{eq:betadist_fit} and \ref{eq:circbetadist_fit}). Second, in most TDEs the low-mass star population is expected to be enhanced as compared to their intrinsic weight, unless when $M_{\rm BH} \gtrsim 10^7\,M_\odot$ for which low-mass stars directly plunge (Fig.\ref{fig:mstardist}).  
\end{itemize}

With the next-generation all-sky transient surveys such as the Vera Rubin Observatory, the TDE sample size will increase by tens of times, making it possible to use TDEs as probes of the demographics of massive black holes and stars in the centers of galaxies.  In this work, using first principles, we have demonstrated that it is important to understand how physical processes in TDEs, such as how promptly the debris forms a disk, can significantly impact the distributions of $M_{\rm BH}$ and $m_\star$ obtained from TDE observations.  

It is important to stress that our study aims at providing one important link narrowing the gap between the intrinsic and the observed TDE rates and demographics. There can be other factors are affecting the observed TDE $M_{\rm BH}$ distribution, such as dust obscuration/reddening and the sensitivity and bias of the instruments. Also, the chance of detecting a TDE necessarily depends on its luminosity (in monochromatic bands), and more studies of TDE emission processes are required for understanding how the luminosity links to $M_{\rm BH}$ and other parameters.  Moreover, we acknowledge that the low-mass end of the BHMF still has large uncertainties, which directly links to the rates of TDEs around low-mass SMBHs.  Eventually, comprehensive forward modeling including all these factors will be needed to bring the theory and observation together.

\acknowledgments

\noindent We thank Alex Dittmann, Tiara Hung, Cole Miller, Brenna Mockler, and Sjoert van Velzen for useful comments and discussions. The authors acknowledge the support from the Hong Kong Research Grants Council (HKU27305119, HKU17304821) and the National Natural Science Foundation of China (HKU12122309).

\bibliography{main}

\begin{thebibliography}{}
\expandafter\ifx\csname natexlab\endcsname\relax\def\natexlab#1{#1}\fi
\providecommand{\url}[1]{\href{#1}{#1}}

\bibitem[{{Auchettl} {et~al.}(2017){Auchettl}, {Guillochon}, \&
  {Ramirez-Ruiz}}]{2017ApJ...838..149A}
{Auchettl}, K., {Guillochon}, J., \& {Ramirez-Ruiz}, E. 2017, \apj, 838, 149

\bibitem[{{Bonnerot} \& {Lu}(2020)}]{2020MNRAS.495.1374B}
{Bonnerot}, C., \& {Lu}, W. 2020, \mnras, 495, 1374

\bibitem[{{Bonnerot} \& {Stone}(2021)}]{2021SSRv..217...16B}
{Bonnerot}, C., \& {Stone}, N.~C. 2021, \ssr, 217, 16

\bibitem[{{Dai} {et~al.}(2021){Dai}, {Lodato}, \&
  {Cheng}}]{2021SSRv..217...12D}
{Dai}, J.~L., {Lodato}, G., \& {Cheng}, R. 2021, \ssr, 217, 12

\bibitem[{{Dai} {et~al.}(2013){Dai}, {Escala}, \&
  {Coppi}}]{2013ApJ...775L...9D}
{Dai}, L., {Escala}, A., \& {Coppi}, P. 2013, \apjl, 775, L9

\bibitem[{{Dai} {et~al.}(2015){Dai}, {McKinney}, \&
  {Miller}}]{2015ApJ...812L..39D}
{Dai}, L., {McKinney}, J.~C., \& {Miller}, M.~C. 2015, \apjl, 812, L39

\bibitem[{{French} {et~al.}(2020){French}, {Wevers}, {Law-Smith}, {Graur}, \&
  {Zabludoff}}]{2020SSRv..216...32F}
{French}, K.~D., {Wevers}, T., {Law-Smith}, J., {Graur}, O., \& {Zabludoff},
  A.~I. 2020, \ssr, 216, 32

\bibitem[{{Gallo} \& {Sesana}(2019)}]{2019ApJ...883L..18G}
{Gallo}, E., \& {Sesana}, A. 2019, \apjl, 883, L18

\bibitem[{{Gezari}(2021)}]{2021arXiv210414580G}
{Gezari}, S. 2021, arXiv e-prints, arXiv:2104.14580

\bibitem[{{Graur} {et~al.}(2018){Graur}, {French}, {Zahid}, {Guillochon},
  {Mandel}, {Auchettl}, \& {Zabludoff}}]{2018ApJ...853...39G}
{Graur}, O., {French}, K.~D., {Zahid}, H.~J., {et~al.} 2018, \apj, 853, 39

\bibitem[{{Guillochon} \& {Ramirez-Ruiz}(2013)}]{2013ApJ...767...25G}
{Guillochon}, J., \& {Ramirez-Ruiz}, E. 2013, \apj, 767, 25

\bibitem[{{Guillochon} \& {Ramirez-Ruiz}(2015)}]{2015ApJ...809..166G}
---. 2015, \apj, 809, 166

\bibitem[{{Hung} {et~al.}(2017){Hung}, {Gezari}, {Blagorodnova}, {Roth},
  {Cenko}, {Kulkarni}, {Horesh}, {Arcavi}, {McCully}, {Yan}, {Lunnan},
  {Fremling}, {Cao}, {Nugent}, \& {Wozniak}}]{2017ApJ...842...29H}
{Hung}, T., {Gezari}, S., {Blagorodnova}, N., {et~al.} 2017, \apj, 842, 29

\bibitem[{{Kesden}(2012)}]{2012PhRvD..85b4037K}
{Kesden}, M. 2012, \prd, 85, 024037

\bibitem[{{Kippenhahn} \& {Weigert}(1990)}]{1990sse..book.....K}
{Kippenhahn}, R., \& {Weigert}, A. 1990, {Stellar Structure and Evolution}

\bibitem[{{Kochanek}(1994)}]{1994ApJ...422..508K}
{Kochanek}, C.~S. 1994, \apj, 422, 508

\bibitem[{{Kochanek}(2016)}]{2016MNRAS.461..371K}
---. 2016, \mnras, 461, 371

\bibitem[{{Komossa} {et~al.}(2008){Komossa}, {Zhou}, {Wang}, {Ajello}, {Ge},
  {Greiner}, {Lu}, {Salvato}, {Saxton}, {Shan}, {Xu}, \&
  {Yuan}}]{2008ApJ...678L..13K}
{Komossa}, S., {Zhou}, H., {Wang}, T., {et~al.} 2008, \apjl, 678, L13

\bibitem[{{Kroupa}(2001)}]{2001MNRAS.322..231K}
{Kroupa}, P. 2001, \mnras, 322, 231

\bibitem[{{Law-Smith} {et~al.}(2019){Law-Smith}, {Guillochon}, \&
  {Ramirez-Ruiz}}]{2019ApJ...882L..25L}
{Law-Smith}, J., {Guillochon}, J., \& {Ramirez-Ruiz}, E. 2019, \apjl, 882, L25

\bibitem[{{Law-Smith} {et~al.}(2020){Law-Smith}, {Coulter}, {Guillochon},
  {Mockler}, \& {Ramirez-Ruiz}}]{2020ApJ...905..141L}
{Law-Smith}, J. A.~P., {Coulter}, D.~A., {Guillochon}, J., {Mockler}, B., \&
  {Ramirez-Ruiz}, E. 2020, \apj, 905, 141

\bibitem[{{Liptai} {et~al.}(2019){Liptai}, {Price}, {Mandel}, \&
  {Lodato}}]{2019arXiv191010154L}
{Liptai}, D., {Price}, D.~J., {Mandel}, I., \& {Lodato}, G. 2019, arXiv
  e-prints, arXiv:1910.10154

\bibitem[{{Lu} \& {Bonnerot}(2020)}]{2020MNRAS.492..686L}
{Lu}, W., \& {Bonnerot}, C. 2020, \mnras, 492, 686

\bibitem[{{Magorrian} \& {Tremaine}(1999)}]{1999MNRAS.309..447M}
{Magorrian}, J., \& {Tremaine}, S. 1999, \mnras, 309, 447

\bibitem[{{Merritt}(2013)}]{2013CQGra..30x4005M}
{Merritt}, D. 2013, Classical and Quantum Gravity, 30, 244005

\bibitem[{{Merritt} \& {Ferrarese}(2001)}]{2001ASPC..249..335M}
{Merritt}, D., \& {Ferrarese}, L. 2001, in Astronomical Society of the Pacific
  Conference Series, Vol. 249, The Central Kiloparsec of Starbursts and AGN:
  The La Palma Connection, ed. J.~H. {Knapen}, J.~E. {Beckman}, I.~{Shlosman},
  \& T.~J. {Mahoney}, 335

\bibitem[{{Mockler} {et~al.}(2019){Mockler}, {Guillochon}, \&
  {Ramirez-Ruiz}}]{2019ApJ...872..151M}
{Mockler}, B., {Guillochon}, J., \& {Ramirez-Ruiz}, E. 2019, \apj, 872, 151

\bibitem[{{Pfister} {et~al.}(2019){Pfister}, {Bar-Or}, {Volonteri}, {Dubois},
  \& {Capelo}}]{2019MNRAS.488L..29P}
{Pfister}, H., {Bar-Or}, B., {Volonteri}, M., {Dubois}, Y., \& {Capelo}, P.~R.
  2019, \mnras, 488, L29

\bibitem[{{Pfister} {et~al.}(2021){Pfister}, {Toscani}, {Wong}, {Lixin Dai},
  {Lodato}, \& {Rossi}}]{2021arXiv210305883P}
{Pfister}, H., {Toscani}, M., {Wong}, T. H.~T., {et~al.} 2021, arXiv e-prints,
  arXiv:2103.05883

\bibitem[{{Pfister} {et~al.}(2020){Pfister}, {Volonteri}, {Dai}, \&
  {Colpi}}]{2020MNRAS.497.2276P}
{Pfister}, H., {Volonteri}, M., {Dai}, J.~L., \& {Colpi}, M. 2020, \mnras, 497,
  2276

\bibitem[{{Piran} {et~al.}(2015){Piran}, {Svirski}, {Krolik}, {Cheng}, \&
  {Shiokawa}}]{2015ApJ...806..164P}
{Piran}, T., {Svirski}, G., {Krolik}, J., {Cheng}, R.~M., \& {Shiokawa}, H.
  2015, \apj, 806, 164

\bibitem[{{Rees}(1988)}]{1988Natur.333..523R}
{Rees}, M.~J. 1988, \nat, 333, 523

\bibitem[{{Reines} \& {Volonteri}(2015)}]{2015ApJ...813...82R}
{Reines}, A.~E., \& {Volonteri}, M. 2015, \apj, 813, 82

\bibitem[{{Roth} {et~al.}(2018){Roth}, {Mushotzky}, {Gezari}, \& {van
  Velzen}}]{2018AAS...23140503R}
{Roth}, N., {Mushotzky}, R., {Gezari}, S., \& {van Velzen}, S. 2018, in
  American Astronomical Society Meeting Abstracts, Vol. 231, American
  Astronomical Society Meeting Abstracts \#231, 405.03

\bibitem[{{Roth} {et~al.}(2020){Roth}, {Rossi}, {Krolik}, {Piran}, {Mockler},
  \& {Kasen}}]{2020SSRv..216..114R}
{Roth}, N., {Rossi}, E.~M., {Krolik}, J., {et~al.} 2020, \ssr, 216, 114

\bibitem[{{Roth} {et~al.}(2021){Roth}, {van Velzen}, {Cenko}, \&
  {Mushotzky}}]{2021ApJ...910...93R}
{Roth}, N., {van Velzen}, S., {Cenko}, S.~B., \& {Mushotzky}, R.~F. 2021, \apj,
  910, 93

\bibitem[{{Ryu} {et~al.}(2020{\natexlab{a}}){Ryu}, {Krolik}, \&
  {Piran}}]{2020ApJ...904...73R}
{Ryu}, T., {Krolik}, J., \& {Piran}, T. 2020{\natexlab{a}}, \apj, 904, 73

\bibitem[{{Ryu} {et~al.}(2020{\natexlab{b}}){Ryu}, {Krolik}, {Piran}, \&
  {Noble}}]{2020ApJ...904...98R}
{Ryu}, T., {Krolik}, J., {Piran}, T., \& {Noble}, S.~C. 2020{\natexlab{b}},
  \apj, 904, 98

\bibitem[{{Saxton} {et~al.}(2020){Saxton}, {Komossa}, {Auchettl}, \&
  {Jonker}}]{2020SSRv..216...85S}
{Saxton}, R., {Komossa}, S., {Auchettl}, K., \& {Jonker}, P.~G. 2020, \ssr,
  216, 85

\bibitem[{{Shankar} {et~al.}(2016){Shankar}, {Bernardi}, {Sheth}, {Ferrarese},
  {Graham}, {Savorgnan}, {Allevato}, {Marconi}, {L{\"a}sker}, \&
  {Lapi}}]{2016MNRAS.460.3119S}
{Shankar}, F., {Bernardi}, M., {Sheth}, R.~K., {et~al.} 2016, \mnras, 460, 3119

\bibitem[{{Shiokawa} {et~al.}(2015){Shiokawa}, {Krolik}, {Cheng}, {Piran}, \&
  {Noble}}]{2015ApJ...804...85S}
{Shiokawa}, H., {Krolik}, J.~H., {Cheng}, R.~M., {Piran}, T., \& {Noble}, S.~C.
  2015, \apj, 804, 85

\bibitem[{{Stone} \& {Metzger}(2016)}]{2016MNRAS.455..859S}
{Stone}, N.~C., \& {Metzger}, B.~D. 2016, \mnras, 455, 859

\bibitem[{{Stone} {et~al.}(2020){Stone}, {Vasiliev}, {Kesden}, {Rossi},
  {Perets}, \& {Amaro-Seoane}}]{2020SSRv..216...35S}
{Stone}, N.~C., {Vasiliev}, E., {Kesden}, M., {et~al.} 2020, \ssr, 216, 35

\bibitem[{{Strubbe}(2011)}]{2011PhDT.......385S}
{Strubbe}, L.~E. 2011, PhD thesis, University of California, Berkeley

\bibitem[{{Ulmer}(1998)}]{1998AIPC..431..141U}
{Ulmer}, A. 1998, in American Institute of Physics Conference Series, Vol. 431,
  Accretion processes in Astrophysical Systems: Some like it hot! - eigth
  AstroPhysics Conference, ed. S.~S. {Holt} \& T.~R. {Kallman}, 141--144

\bibitem[{{van Velzen}(2018)}]{2018ApJ...852...72V}
{van Velzen}, S. 2018, \apj, 852, 72

\bibitem[{{van Velzen} {et~al.}(2020){van Velzen}, {Holoien}, {Onori}, {Hung},
  \& {Arcavi}}]{2020SSRv..216..124V}
{van Velzen}, S., {Holoien}, T. W.~S., {Onori}, F., {Hung}, T., \& {Arcavi}, I.
  2020, \ssr, 216, 124

\bibitem[{{van Velzen} {et~al.}(2021){van Velzen}, {Gezari}, {Hammerstein},
  {Roth}, {Frederick}, {Ward}, {Hung}, {Cenko}, {Stein}, {Perley}, {Taggart},
  {Foley}, {Sollerman}, {Blagorodnova}, {Andreoni}, {Bellm}, {Brinnel}, {De},
  {Dekany}, {Feeney}, {Fremling}, {Giomi}, {Golkhou}, {Graham}, {Ho},
  {Kasliwal}, {Kilpatrick}, {Kulkarni}, {Kupfer}, {Laher}, {Mahabal}, {Masci},
  {Miller}, {Nordin}, {Riddle}, {Rusholme}, {van Santen}, {Sharma}, {Shupe}, \&
  {Soumagnac}}]{2021ApJ...908....4V}
{van Velzen}, S., {Gezari}, S., {Hammerstein}, E., {et~al.} 2021, \apj, 908, 4

\bibitem[{{Wang} \& {Merritt}(2004)}]{2004ApJ...600..149W}
{Wang}, J., \& {Merritt}, D. 2004, \apj, 600, 149

\bibitem[{{Wevers} {et~al.}(2017){Wevers}, {van Velzen}, {Jonker}, {Stone},
  {Hung}, {Onori}, {Gezari}, \& {Blagorodnova}}]{2017MNRAS.471.1694W}
{Wevers}, T., {van Velzen}, S., {Jonker}, P.~G., {et~al.} 2017, \mnras, 471,
  1694

\bibitem[{{Wevers} {et~al.}(2019){Wevers}, {Stone}, {van Velzen}, {Jonker},
  {Hung}, {Auchettl}, {Gezari}, {Onori}, {Mata S{\'a}nchez},
  {Kostrzewa-Rutkowska}, \& {Casares}}]{2019MNRAS.487.4136W}
{Wevers}, T., {Stone}, N.~C., {van Velzen}, S., {et~al.} 2019, \mnras, 487,
  4136

\end{thebibliography}
\bibliographystyle{aasjournal}

\appendix
\counterwithin{figure}{section}
\counterwithin{table}{section}
\section{TDE Host Black Hole Masses \label{appsec:TDEdist}}
\noindent In this appendix, we present the criteria used to select the likely TDEs among the many other TDE candidates as our observed sample, and further elaborate on the properties of both the host black hole mass and luminosity distributions we find.

\subsection{TDE Selection and Black Hole Mass Calculation \label{appsubsec:MBHdist}}
\noindent Among a total of a hundred or so detected TDE candidates, we select 65 TDEs to include in our analysis.  The chosen TDE candidates are listed in Table \ref{TDEtable} with some of their host galaxy properties.  Detailed criteria for the selection of TDE candidates are as follows. 

\begin{itemize}
    \item Categorized as a possible/likely/confirmed TDE in at least \emph{two} of the following literature \citep{2017ApJ...838..149A, 2017MNRAS.471.1694W, 2019MNRAS.487.4136W, 2020SSRv..216...32F, 2021ApJ...908....4V}
    \item All TDEs in Table 1 of \citet{2021arXiv210414580G}
    \item All ZTF TDEs  \citep{2021ApJ...908....4V}
\end{itemize}
Also, we intentionally exclude all likely partial TDEs, e.g.,\ iPTF-16fnl and PS1-11af, from our sample, since our calculation only applies to fully disrupted stars.  

We also obtain the TDE host black hole mass with two different methods. We either derive their $M_{\rm BH}$ all using the $M_{\rm BH}-\sigma$ relation \citep{2001ASPC..249..335M}:
\begin{equation}
    \log_{10}\left(\frac{M_{{\rm BH},\sigma}}{M_\odot}\right) = \log_{10}\left[10^8\cdot\left(1.48\pm0.24\right)\left(\frac{\sigma}{200\,\text{km s}^{-1}}\right)^{4.65\pm0.48}\right] \label{eq:Msigma}
\end{equation}
or using the $M_{\rm BH}-M_{\text{gal}}$ relation \citep{2015ApJ...813...82R}:
\begin{equation}
    \log_{10}\left(\frac{M_{{\rm BH},\text{gal}}}{M_\odot}\right) = \left(7.45\pm0.08\right) + \left(1.05\pm0.11\right)\log_{10}\left(\frac{M_{\text{gal}}}{10^{11}\,M_\odot}\right) \label{eq:Mgal}
\end{equation}
The derived black hole masses as well as other quantities such as the peak luminosity (assuming blackbody emission) are listed in Table \ref{TDEtable}. The error bar of each derived $M_{\rm BH}$ is computed by modeling a Gaussian distribution using the measurement errors of the velocity dispersion $\sigma$ and the total galaxy mass $M_{\rm gal}$, with the curve-fitting $1\sigma$ error of Eq.\ref{eq:Msigma} and Eq.\ref{eq:Mgal}.  The Gaussian distributions peak at the mean value $\log_{10}\bar{M}_{\rm BH}$ with standard deviation $s=\left(\log_{10}M_{\rm BH}^{\text{up}}-\log_{10}M_{\rm BH}^{\text{low}}\right)/2$.  

\begin{figure*}
\centering
\includegraphics[width=1.00\linewidth]{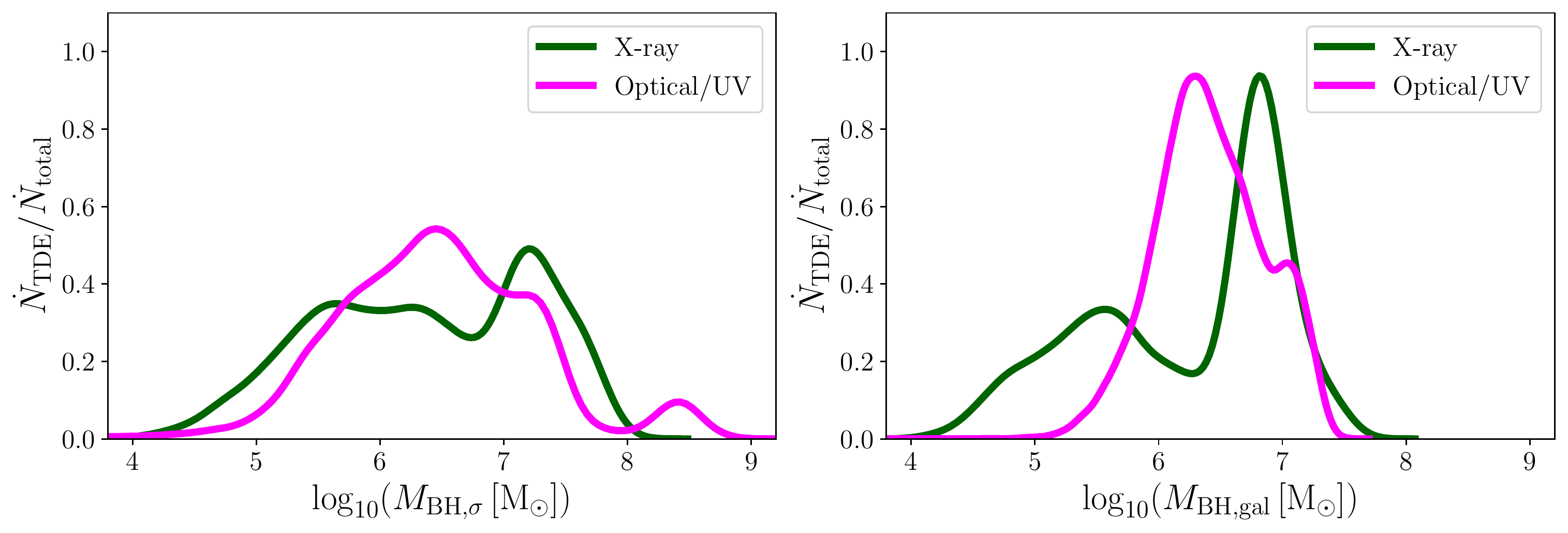}
\caption{\textbf{TDE host $M_{\rm BH}$ distribution for X-ray strong or optical/UV strong TDEs}.  The two coloured curves in Fig.\ref{fig:MBHdistGG} are further decomposed into two groups: optical/UV strong (magenta) or X-ray strong (green). The optical/UV TDE $M_{\rm BH}$ distribution still peaks somewhere between $10^6-10^7 M_\odot$, while that of the X-ray TDEs exhibits a double peak behavior.  \label{fig:MBHdistXOGG}}
\end{figure*}

We plot the $M_{\rm BH}$ distribution of all TDE samples obtained from either method in Fig.\ref{fig:MBHdistGG}. The overall trends are similar, but the detailed shapes of the two distributions are somewhat different from each other. The main reason for this the that for many TDEs we lack measurements of $\sigma$.  We further separate between the optical/UV strong (magenta) and X-ray strong (green) TDE samples in Fig.\ref{fig:MBHdistXOGG}. It can be clearly seen that the X-ray strong TDEs have a bimodal distribution in $M_{\rm BH}$, but we believe that this effect is primarily due to the limited sample size of this TDE category.  Thus, shifting of the distribution, perhaps reducing to a single peak, is highly likely when more are observed.  Another possibility is related to the TDE emission mechanism in different wavebands which deserves extensive study.  
\newpage

\startlongtable
\begin{deluxetable*}{llccccc}
\centering
\tablecaption{Chosen TDE Candidates with Their Host Galaxy Properties\label{TDEtable}}
\tablewidth{700pt}
\tablehead{
\colhead{Name} & \colhead{Redshift} & 
\colhead{$\sigma$ [km s$^{-1}$]} & \colhead{$\log_{10}\left(\displaystyle\frac{M_{{\rm BH},\sigma}}{M_\odot}\right)$} & \colhead{$\log_{10}\left(\displaystyle\frac{M_{\text{gal}}}{M_\odot}\right)$} & 
\colhead{$\log_{10}\left(\displaystyle\frac{M_{{\rm BH},\text{gal}}}{M_\odot}\right)$} & \colhead{$\log_{10}\left(L_{\text{bb},\text{peak}}\left[\text{erg s}^{-1}\right]\right)$}}
\startdata
\textit{X-ray TDEs} & & & & & & \\
\tableline
2MASX J0249 & 0.0186 & $43^{+4}_{-4}$ $^{\text{f}}$ & $5.07^{+0.55}_{-0.62}$ & $9.1^{+...}_{-...}$ $^{\text{e}}$ & $5.46^{+...}_{-...}$ & \nodata \\
2XMMi J1847-63 & 0.0353 & \nodata & \nodata & \nodata & \nodata & $42.82^{+...}_{-...}$ $^{\text{h}}$ \\
3XMM J1500 & 0.1454 & $59^{+3}_{-3}$ $^{\text{e}}$ & $5.71^{+0.41}_{-0.45}$ & $9.3^{+...}_{-...}$ $^{\text{e}}$ & $5.67^{+...}_{-...}$ & $43.08^{+...}_{-...}$ $^{\text{h}}$ \\
3XMM J1521+0749 & 0.179 & $58^{+2}_{-2}$ $^{\text{f}}$ & $5.67^{+0.39}_{-0.41}$ & $10.17^{+0.11}_{-0.20}$ $^{\text{c}}$ & $6.58^{+0.27}_{-0.40}$ & $43.51^{+...}_{-...}$ $^{\text{h}}$ \\
ASASSN-14li & 0.0206 & $81^{+2}_{-2}$ $^{\text{b}}$ & $6.35^{+0.30}_{-0.32}$ & $9.71^{+0.05}_{-0.10}$ $^{\text{g}}$ & $6.10^{+0.26}_{-0.34}$ & $43.66^{+0.02}_{-0.02}$ $^{\text{g}}$ \\
LEDA 095953 & 0.0366 & \nodata & \nodata & \nodata & \nodata & \nodata \\
NGC 5905 & 0.011 & $97^{+5}_{-5}$ $^{\text{f}}$ & $6.71^{+0.31}_{-0.35}$ & $10.83^{+0.22}_{-0.06}$ $^{\text{c}}$ & $7.27^{+0.31}_{-0.17}$ & $40.94^{+...}_{-...}$ $^{\text{h}}$ \\
RBS 1032 & 0.026 & $49^{+7}_{-7}$ $^{\text{f}}$ & $5.33^{+0.60}_{-0.71}$ & $9.19^{+0.15}_{-0.16}$ $^{\text{c}}$ & $5.55^{+0.42}_{-0.47}$ & $41.70^{+...}_{-...}$ $^{\text{h}}$ \\
RX J1242-1119-A & 0.05 & \nodata & \nodata & $10.3^{+...}_{-...}$ $^{\text{e}}$ & $6.72^{+...}_{-...}$ & $42.60^{+...}_{-...}$ $^{\text{h}}$ \\
RX J1420+5334-A & 0.147 & $131^{+13}_{-13}$ $^{\text{f}}$ & $7.32^{+0.33}_{-0.40}$ & $10.53^{+0.07}_{-0.07}$ $^{\text{c}}$ & $6.96^{+0.19}_{-0.22}$ & $43.38^{+...}_{-...}$ $^{\text{h}}$ \\
RX J1624+7554 & 0.0636 & $155^{+9}_{-9}$ $^{\text{f}}$ & $7.66^{+0.22}_{-0.26}$ & $10.4^{+...}_{-...}$ $^{\text{e}}$ & $6.82^{+...}_{-...}$ & $43.38^{+...}_{-...}$ $^{\text{h}}$ \\
SDSS J0159 & 0.3117 & $124^{+10}_{-10}$ $^{\text{f}}$ & $7.21^{+0.31}_{-0.36}$ & $10.37^{+0.11}_{-0.06}$ $^{\text{c}}$ & $6.79^{+0.25}_{-0.22}$ & \nodata \\
SDSS J1201+3003 & 0.146 & $122^{+4}_{-4}$ $^{\text{f}}$ & $7.17^{+0.23}_{-0.25}$ & $10.61^{+0.08}_{-0.16}$ $^{\text{c}}$ & $7.04^{+0.20}_{-0.31}$ & $45.00^{+...}_{-...}$ $^{\text{h}}$ \\
SDSS J1311-0123 & 0.195 & \nodata & \nodata & $8.7^{+...}_{-...}$ $^{\text{f}}$ & $5.04^{+...}_{-...}$ & $41.74^{+...}_{-...}$ $^{\text{h}}$ \\
SDSS J1323+4827 & 0.0875 & $75^{+4}_{-4}$ $^{\text{f}}$ & $6.19^{+0.36}_{-0.40}$ & $10.38^{+0.06}_{-0.07}$ $^{\text{c}}$ & $6.80^{+0.20}_{-0.23}$ & $44.30^{+...}_{-...}$ $^{\text{h}}$ \\
Swift J1112 & 0.89 & \nodata & \nodata & \nodata & \nodata & \nodata \\
Swift J1644+57 & 0.3543 & \nodata & \nodata & \nodata & \nodata & \nodata \\
Swift J2058 & 1.186 & \nodata & \nodata & \nodata & \nodata & \nodata \\
TDXF J1347-32 & 0.0366 & \nodata & \nodata & \nodata & \nodata & $42.73^{+...}_{-...}$ $^{\text{h}}$ \\
WINGS J1348+26 & 0.0651 & \nodata & \nodata & $8.48^{+...}_{-...}$ $^{\text{s}}$ & $4.80^{+...}_{-...}$ & $41.79^{+...}_{-...}$ $^{\text{h}}$ \\
XMM J0740 & 0.0173 & \nodata & \nodata & \nodata & \nodata & $42.61^{+...}_{-...}$ $^{\text{h}}$ \\
\tableline
\textit{Optical/UV TDEs} & & & & & & \\
\tableline
ASASSN-14ae & 0.044 & $53^{+2}_{-2}$ $^{\text{f}}$ & $5.49^{+0.41}_{-0.44}$ & $9.73^{+0.13}_{-0.13}$ $^{\text{c}}$ & $6.12^{+0.34}_{-0.37}$ & $43.87^{+0.01}_{-0.01}$ $^{\text{g}}$ \\
ASASSN-14li & 0.0206 & $81^{+2}_{-2}$ $^{\text{b}}$ & $6.35^{+0.30}_{-0.32}$ & $9.71^{+0.05}_{-0.10}$ $^{\text{g}}$ & $6.10^{+0.26}_{-0.34}$ & $43.66^{+0.02}_{-0.02}$ $^{\text{g}}$ \\
ASASSN-15lh & 0.233 & $225^{+15}_{-15}$ $^{\text{d}}$ & $8.41^{+0.16}_{-0.21}$ & $10.8^{+...}_{-...}$ $^{\text{d}}$ & $7.24^{+...}_{-...}$ & $45.6^{+...}_{-...}$ $^{\text{d}}$ \\
ASASSN-15oi & 0.048 & $61^{+7}_{-7}$ $^{\text{f}}$ & $5.77^{+0.51}_{-0.60}$ & $10.05^{+0.04}_{-0.04}$ $^{\text{g}}$ & $6.45^{+0.22}_{-0.23}$ & $44.45^{+0.01}_{-0.01}$ $^{\text{g}}$ \\
ASASSN-19dj & 0.022 & \nodata & \nodata & $9.82^{+0.16}_{-0.13}$ $^{\text{i}}$ & $6.21^{+0.36}_{-0.36}$ & $44.50^{+0.02}_{-0.02}$ $^{\text{g}}$ \\
AT2018bsi & 0.051 & \nodata & \nodata & $10.63^{+0.05}_{-0.05}$ $^{\text{i}}$ & $7.06^{+0.17}_{-0.18}$ & $43.87^{+0.08}_{-0.08}$ $^{\text{g}}$ \\
AT2018dyb & 0.018 & $96^{+1}_{-1}$ $^{\text{s}}$ & $6.69^{+0.24}_{-0.25}$ & $9.86^{+0.08}_{-0.15}$ $^{\text{g}}$ & $6.25^{+0.28}_{-0.38}$ & $44.08^{+...}_{-...}$ $^{\text{h}}$ \\
AT2018dyk & 0.037 & $112^{+4}_{-4}$ $^{\text{f}}$ & $7.00^{+0.25}_{-0.28}$ & $10.6^{+...}_{-...}$ $^{\text{f}}$ & $7.03^{+...}_{-...}$ & \nodata \\
AT2018hco & 0.088 & \nodata & \nodata & $9.95^{+0.12}_{-0.16}$ $^{\text{i}}$ & $6.35^{+0.31}_{-0.38}$ & $44.25^{+0.04}_{-0.04}$ $^{\text{g}}$ \\
AT2018hyz & 0.0458 & $60^{+5}_{-5}$ $^{\text{f}}$ & $5.74^{+0.46}_{-0.52}$ & $9.84^{+0.09}_{-0.14}$ $^{\text{i}}$ & $6.23^{+0.29}_{-0.37}$ & $44.10^{+0.01}_{-0.01}$ $^{\text{g}}$ \\
AT2018fyk & 0.059 & \nodata & \nodata & $10.58^{+0.12}_{-0.21}$ $^{\text{g}}$ & $7.01^{+0.24}_{-0.37}$ & $44.48^{+...}_{-...}$ $^{\text{h}}$ \\
AT2018iih & 0.212 & \nodata & \nodata & $10.63^{+0.18}_{-0.14}$ $^{\text{i}}$ & $7.06^{+0.29}_{-0.28}$ & $44.62^{+0.04}_{-0.04}$ $^{\text{g}}$ \\
AT2018lna & 0.091 & \nodata & \nodata & $9.49^{+0.11}_{-0.12}$ $^{\text{i}}$ & $5.86^{+0.35}_{-0.38}$ & $44.56^{+0.06}_{-0.06}$ $^{\text{g}}$ \\
AT2018lni & 0.138 & \nodata & \nodata & $10.00^{+0.09}_{-0.14}$ $^{\text{i}}$ & $6.40^{+0.27}_{-0.35}$ & $44.21^{+0.29}_{-0.17}$ $^{\text{g}}$ \\
AT2019ahk & 0.0262 & \nodata & \nodata & $9.72^{+0.09}_{-0.10}$ $^{\text{g}}$ & $6.11^{+0.30}_{-0.34}$ & $44.08^{+...}_{-...}$ $^{\text{h}}$ \\
AT2019bhf & 0.1206 & \nodata & \nodata & $10.25^{+0.14}_{-0.12}$ $^{\text{i}}$ & $6.66^{+0.30}_{-0.30}$ & $43.91^{+0.04}_{-0.05}$ $^{\text{g}}$ \\
AT2019cho & 0.193 & \nodata & \nodata & $10.20^{+0.11}_{-0.14}$ $^{\text{i}}$ & $6.61^{+0.27}_{-0.33}$ & $43.98^{+0.01}_{-0.01}$ $^{\text{g}}$ \\
AT2019dsg & 0.0512 & \nodata & \nodata & $10.46^{+0.11}_{-0.19}$ $^{\text{i}}$ & $6.88^{+0.25}_{-0.36}$ & $44.26^{+0.04}_{-0.05}$ $^{\text{g}}$ \\
AT2019ehz & 0.074 & \nodata & \nodata & $9.74^{+0.08}_{-0.09}$ $^{\text{i}}$ & $6.13^{+0.29}_{-0.33}$ & $44.03^{+0.01}_{-0.02}$ $^{\text{g}}$ \\
AT2019eve & 0.064 & \nodata & \nodata & $9.31^{+0.10}_{-0.15}$ $^{\text{i}}$ & $5.68^{+0.36}_{-0.44}$ & $43.14^{+0.02}_{-0.03}$ $^{\text{g}}$ \\
AT2019lwu & 0.117 & \nodata & \nodata & $9.86^{+0.09}_{-0.13}$ $^{\text{i}}$ & $6.25^{+0.29}_{-0.35}$ & $43.60^{+0.03}_{-0.04}$ $^{\text{g}}$ \\
AT2019meg & 0.152 & \nodata & \nodata & $9.70^{+0.15}_{-0.08}$ $^{\text{i}}$ & $6.09^{+0.36}_{-0.32}$ & $44.36^{+0.03}_{-0.04}$ $^{\text{g}}$ \\
AT2019mha & 0.148 & \nodata & \nodata & $10.07^{+0.10}_{-0.18}$ $^{\text{i}}$ & $6.47^{+0.28}_{-0.39}$ & $44.05^{+0.06}_{-0.05}$ $^{\text{g}}$ \\
AT2019qiz & 0.0151 & \nodata & \nodata & $10.01^{+0.09}_{-0.13}$ $^{\text{i}}$ & $6.41^{+0.27}_{-0.34}$ & $43.44^{+0.01}_{-0.01}$ $^{\text{g}}$ \\
F01004 & 0.1178 & $132^{+29}_{-29}$ $^{\text{f}}$ & $7.33^{+0.51}_{-0.72}$ & $9.8^{+...}_{-...}$ $^{\text{f}}$ & $6.19^{+...}_{-...}$ & \nodata \\
GALEX-D1-9 & 0.326 & $89^{+4}_{-4}$ $^{\text{e}}$ & $6.54^{+0.31}_{-0.35}$ & $10.3^{+...}_{-...}$ $^{\text{d}}$ & $6.72^{+...}_{-...}$ & $44.1^{+...}_{-...}$ $^{\text{d}}$ \\
GALEX-D23-H1 & 0.1855 & $84^{+4}_{-4}$ $^{\text{f}}$ & $6.42^{+0.33}_{-0.37}$ & $10.08^{+0.15}_{-0.07}$ $^{\text{c}}$ & $6.48^{+0.33}_{-0.26}$ & $43.9^{+...}_{-...}$ $^{\text{d}}$ \\
GALEX-D3-13 & 0.3698 & $133^{+6}_{-6}$ $^{\text{f}}$ & $7.35^{+0.23}_{-0.27}$ & $10.7^{+...}_{-...}$ $^{\text{c}}$ & $7.14^{+...}_{-...}$ & $44.3^{+...}_{-...}$ $^{\text{d}}$ \\
iPTF-15af & 0.079 & $106^{+2}_{-2}$ $^{\text{f}}$ & $6.89^{+0.23}_{-0.25}$ & $10.31^{+0.08}_{-0.10}$ $^{\text{c}}$ & $6.73^{+0.23}_{-0.28}$ & $44.10^{+0.10}_{-0.08}$ $^{\text{g}}$ \\
iPTF-16axa & 0.108 & $82^{+3}_{-3}$ $^{\text{f}}$ & $6.37^{+0.32}_{-0.35}$ & $10.25^{+0.05}_{-0.08}$ $^{\text{c}}$ & $6.66^{+0.21}_{-0.25}$ & $43.82^{+0.03}_{-0.02}$ $^{\text{g}}$ \\
OGLE16aaa & 0.1655 & \nodata & \nodata & $10.43^{+0.09}_{-0.11}$ $^{\text{g}}$ & $6.85^{+0.23}_{-0.27}$ & $44.22^{+...}_{-...}$ $^{\text{h}}$ \\
PS1-10jh & 0.1696 & $65^{+3}_{-3}$ $^{\text{f}}$ & $5.90^{+0.38}_{-0.42}$ & $9.63^{+0.10}_{-0.13}$ $^{\text{g}}$ & $6.01^{+0.33}_{-0.38}$ & $44.47^{+0.07}_{-0.07}$ $^{\text{g}}$ \\
PS16dtm & 0.0804 & $45^{+13}_{-13}$ $^{\text{f}}$ & $5.16^{+0.84}_{-1.15}$ & $9.77^{+0.11}_{-0.13}$ $^{\text{c}}$ & $6.16^{+0.32}_{-0.37}$ & \nodata \\
PS17dhz & 0.1089 & \nodata & \nodata & $9.47^{+0.11}_{-0.13}$ $^{\text{g}}$ & $5.84^{+0.36}_{-0.40}$ & $43.82^{+...}_{-...}$ $^{\text{h}}$ \\
PS18kh & 0.075 & \nodata & \nodata & $9.95^{+0.12}_{-0.24}$ $^{\text{i}}$ & $6.35^{+0.31}_{-0.48}$ & $43.78^{+0.02}_{-0.02}$ $^{\text{g}}$ \\
PTF-09axc & 0.115 & $60^{+4}_{-4}$ $^{\text{f}}$ & $5.74^{+0.43}_{-0.48}$ & $9.84^{+0.06}_{-0.09}$ $^{\text{c}}$ & $6.23^{+0.27}_{-0.31}$ & $43.46^{+0.03}_{-0.02}$ $^{\text{g}}$ \\
PTF-09djl & 0.184 & $64^{+7}_{-7}$ $^{\text{f}}$ & $5.87^{+0.49}_{-0.57}$ & $9.91^{+0.13}_{-0.17}$ $^{\text{c}}$ & $6.31^{+0.32}_{-0.40}$ & $44.42^{+0.04}_{-0.04}$ $^{\text{g}}$ \\
PTF-09ge & 0.064 & $82^{+2}_{-2}$ $^{\text{f}}$ & $6.37^{+0.30}_{-0.32}$ & $9.87^{+0.13}_{-0.17}$ $^{\text{c}}$ & $6.26^{+0.33}_{-0.40}$ & $44.04^{+0.01}_{-0.01}$ $^{\text{g}}$ \\
PTF-10iya & 0.224 & \nodata & \nodata & $9.3^{+...}_{-...}$ $^{\text{f}}$ & $5.67^{+...}_{-...}$ & \nodata \\
SDSS-TDE1 & 0.136 & $126^{+7}_{-7}$ $^{\text{b}}$ & $7.24^{+0.26}_{-0.30}$ & $10.08^{+0.08}_{-0.12}$ $^{\text{c}}$ & $6.48^{+0.26}_{-0.32}$ & $43.5^{+...}_{-...}$ $^{\text{d}}$ \\
SDSS-TDE2 & 0.2515 & \nodata & \nodata & $10.59^{+0.17}_{-0.10}$ $^{\text{g}}$ & $7.02^{+0.28}_{-0.24}$ & $44.54^{+0.08}_{-0.06}$ $^{\text{g}}$ \\
SDSS J0748+4712 & 0.0615 & $126^{+7}_{-7}$ $^{\text{f}}$ & $7.24^{+0.26}_{-0.30}$ & $10.18^{+0.06}_{-0.09}$ $^{\text{c}}$ & $6.59^{+0.23}_{-0.28}$ & \nodata \\
SDSS J0952+2143 & 0.0789 & $95^{+...}_{-...}$ $^{\text{a}}$ & $6.67^{+...}_{-...}$ & $10.37^{+0.06}_{-0.07}$ $^{\text{c}}$ & $6.79^{+0.20}_{-0.23}$ & \nodata \\
SDSS J1342+0530 & 0.0366 & $72^{+6}_{-6}$ $^{\text{f}}$ & $6.11^{+0.42}_{-0.48}$ & $9.64^{+0.23}_{-0.07}$ $^{\text{c}}$ & $6.02^{+0.45}_{-0.31}$ & \nodata \\
SDSS J1350+2916 & 0.0777 & \nodata & \nodata & $9.94^{+0.17}_{-0.20}$ $^{\text{c}}$ & $6.34^{+0.35}_{-0.43}$ & \nodata
\enddata
\tablecomments{The columns are: TDE name, redshift, velocity dispersion $\sigma$, black hole mass $M_{\rm{BH},\sigma}$ derived from the $M_{\rm BH}-\sigma$ relation, galaxy stellar mass $M_{\rm gal}$, black hole mass $M_{\rm{BH},gal}$ derived from $M_{{\rm BH}}-M_{\text{gal}}$ relation, and peak luminosity assuming blackbody emission $L_{\rm bb,peak}$.  The chosen TDEs are separated into either the optical/UV strong or X-ray strong categories, except for ASASSN-14li which is in both categories.  References for the values of $\sigma$, $M_{\text{gal}}$ and $L_{\text{bb,peak}}$: $^{\text{a}}$\citet{2008ApJ...678L..13K},
$^{\text{b}}$\citet{2017MNRAS.471.1694W}, $^{\text{c}}$\citet{2018ApJ...853...39G}, $^{\text{d}}$\citet{2018ApJ...852...72V}, $^{\text{e}}$\citet{2019MNRAS.487.4136W}, $^{\text{f}}$\citet{2020SSRv..216...32F}, $^{\text{g}}$\citet{2020SSRv..216..124V}, $^{\text{h}}$\citet{2021arXiv210414580G}, $^{\text{i}}$\citet{2021ApJ...908....4V}, and $^{\text{s}}$source discovery papers.  Their 1$\sigma$ error bars of $M_{\rm BH}$ are calculated by modeling each TDE as a Gaussian distribution with the scattering from the $M_{\rm BH}$-scaling relations and the measurement errors from $\sigma$ and $M_{\rm gal}$.}
\end{deluxetable*}

\onecolumngrid
\subsection{TDE Luminosity Distribution \label{appsubsec:Ldist}}
\begin{figure*}
\centering
\includegraphics[width=1.00\linewidth]{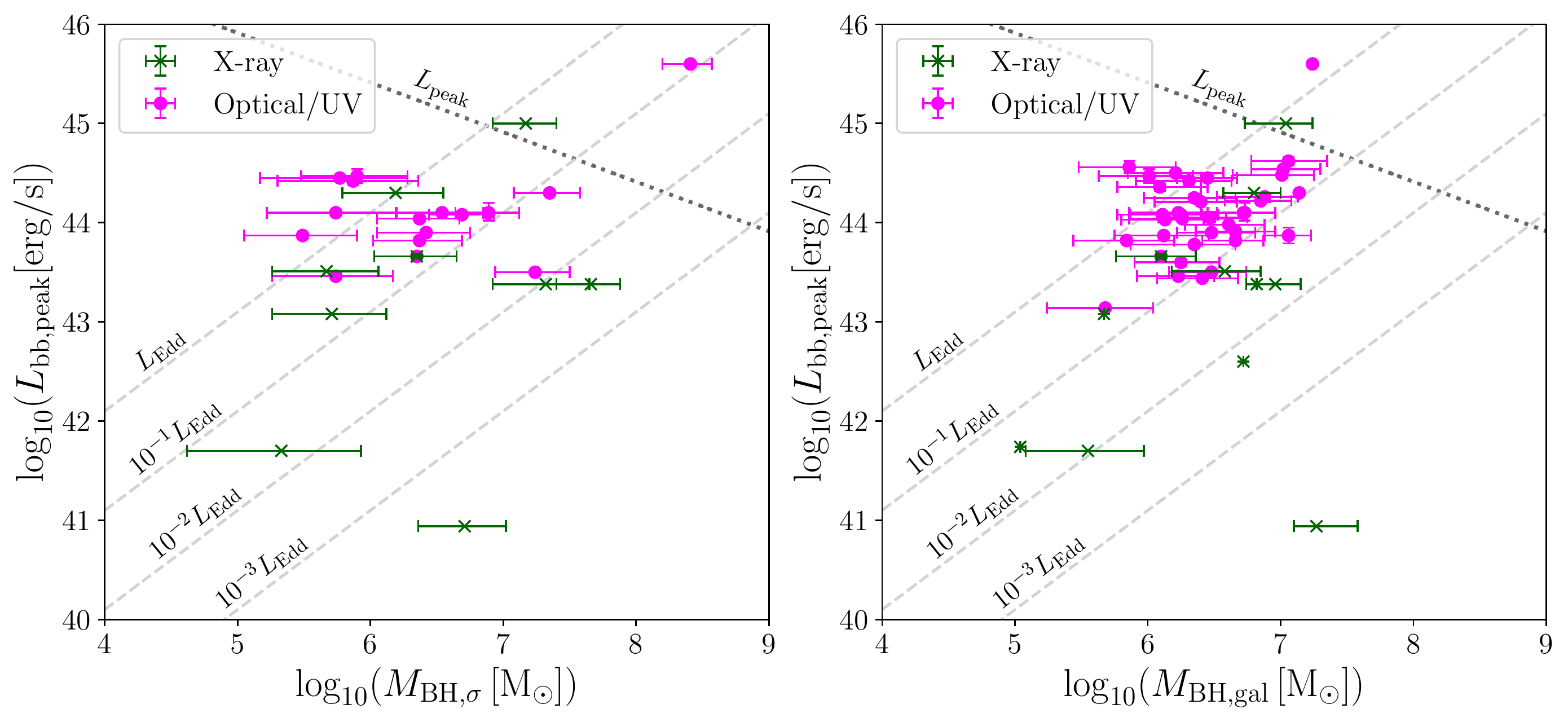}
\caption{\textbf{TDE peak luminosity $L_{\rm bb,peak}$ v.s. $M_{\rm BH}$ from Table \ref{TDEtable}} (modified from Fig.16 \citealt{2017ApJ...842...29H}).  The optical/UV strong TDEs are marked by magenta dots and the X-ray strong by green crosses, with the horizontal line showing the error bar of $M_{\rm BH}$. The dashed lines correspond to four different Eddington ratios, while the dotted line shows the $\dot{M}_{\text{peak}}\propto M_{\rm BH}^{-1/2}$ proportionality relation derived in \citet{2013ApJ...767...25G} Eq.A1 with $m_\star = 1 M_\odot$, $\gamma=4/3$, $\beta=1$, and radiative efficiency $\eta=0.1$.  We find here that some TDEs have peak luminosity exceeding the Eddington limits of their host SMBHs:\ ASASSN-14ae, ASASSN-15lh, ASASSN-15oi, ASASSN-19dj, AT2018hyz, AT2018lna, AT2019meg, PS1-10jh, and SDSS J1323-4827.  Overall, no apparent correlation is found between $L_{\rm bb,peak}$ and $M_{\rm BH}$.  \label{fig:Lbb}}
\end{figure*}

\noindent In Fig.\ref{fig:Lbb} we plot the TDE peak luminosity $L_{\rm bb,peak}$ distribution against $M_{\rm BH}$.  One might expect that the luminosity should generally increase somewhat linearly with $M_{\rm BH}$ if the Eddington limit is the primary factor determining $L_{\rm bb,peak}$ \citep{2016MNRAS.461..371K}, however, here one could barely see any trend of $L_{\rm bb,peak}$ as a function of $M_{\rm BH}$.  We note that the currently sample size is limited, and the range of observed TDE $M_{\rm BH}$ is narrow, which forbid us from drawing any conclusion on whether or how much the Eddington limit is playing a role in determining the luminosity of TDEs.  We also recognize that there are further layer of complications like the flux limits of different instruments \citep{2018ApJ...852...72V} and dust obscuration from host galaxies \citep{2018AAS...23140503R}.  The number of detected TDEs of a given transient survey has a nonlinear dependency on its effective flux limit, and their limits could be used to derive the maximum redshifts each survey are capable of reaching given a peak luminosity.  To stimulate a detection, not only should the event peak luminosity surpasses the survey sensitivity in specific wavebands, but it also has to have significant contrast compared with the flux of its host galaxy.  Even worse, the effects of dust (neutral gas) at optical/UV (X-ray) wavelengths in the host galaxies would obstruct our observation and selection capability.  Hence the observed TDE samples unavoidably suffer from further external observational bias.  

\newpage
\section{Derivations of Energy Dissipation in Stream-stream Collision  \label{appsec:ssinteract}}

\noindent We summarize the results in \citet{2015ApJ...812L..39D} on debris stream self-crossing and disk formation in TDEs. 
The most bound debris has a specific binding energy of
\begin{equation}
    E_{\text{mb}} = \frac{1}{2}v_T^2 - \frac{GM_{\rm BH}}{r_T-r_\star} \approx -\frac{GM_{\rm BH} r_\star}{r_T^2}\,\,\text{for}\,\,r_T\gg r_\star 
\end{equation}
where $v_T$ is the orbital velocity at the tidal radius $r_T$ and $r_\star$ is the stellar radius.  Furthermore, the semi-major axis and eccentricity of this orbit are $a_{\text{mb}} = r_T^2/2r_\star$ and $e_{\text{mb}} = 1-r_p/a_{\text{mb}} \approx 1-2(m_\star/M_{\rm BH})^{1/3}/\beta$ respectively, where $r_p$ is the pericenter distance and $\beta$ is the penetration parameter.  The stellar debris orbits around a Schwarzschild SMBH goes through  apsidal precession. We approximate the process using an instantaneous shift at $r_p$ with an angle of:
\begin{equation}
    \phi = \frac{6\pi GM_{\rm BH}}{c^2a(1-e^2)}
\end{equation}
Using these, the intersection radius and the angle of the incoming and outgoing debris stream are calculated to be 
\begin{equation}
    R_I(M_{\rm BH},\beta,m_\star) = \frac{(1+e_{\text{mb}})r_T}{\beta\left[1-e_{\text{mb}}\cos(\phi/2)\right]}
\end{equation}
\begin{equation}
    \cos\Theta = \frac{1 - 2e_{\text{mb}}\cos(\phi/2) + e_{\text{mb}}^2\cos\phi}{1 - 2e_{\text{mb}}\cos(\phi/2) + e_{\text{mb}}^2}. 
\end{equation}
Assuming totally inelastic collision of the debris streams and using the conservation of momentum, the debris speeds before collision $v_i$ and after collision $v_f$ are related as
\begin{equation}
    v_f = v_i\cos\left(\Theta/2\right) \label{eq:vf}
\end{equation}
where $v_i$ can be calculated using the conservation of energy: 
\begin{equation}
    -\frac{GM_{\rm BH}}{2a_{\text{mb}}} = -\frac{GM_{\rm BH}}{R_I} + \frac{1}{2}v_i^2.  \label{eq:energyconserve}
\end{equation}
We can then obtain the specific energy loss in the first self-crossing for the calculation of disk formation efficiency
\begin{equation}
    \Delta E_{\text{first}}(M_{\rm BH},\beta,m_\star) = \left|\frac{1}{2}v_f^2 - \frac{1}{2}v_i^2\right| = \frac{1}{2}v_i^2\sin^2\left(\Theta/2\right).  \label{appeq:deltaEfirst}
\end{equation}

\section{Supplementary Information on Loss Cone Dynamics \label{appsec:LC}}
\noindent In classical LC calculations, the commonly used variables are defined as: the specific energy $E$, the specific angular momentum $J$, the scaled dimensionless angular momentum $\mathcal{R}=J^2/J_C^2$, where $J_C$ is the specific angular momentum of a circular orbit. The orbital period $P(E)$ and the stellar distribution function $f(E,\mathcal{R})$ can be derived using these variables \citeg{ 2013CQGra..30x4005M, 2016MNRAS.455..859S}.  Considering $f$ outside of the disruption zone $(\mathcal{R}>\mathcal{R}_{\text{LC}})$, where $\mathcal{R}_{\text{LC}}$ is defined as the critical angular momentum at which stars would inevitably by tidally disrupted if their $\mathcal{R}<\mathcal{R}_{\text{LC}}$: 
\begin{equation}
    \lim_{E\ll GM_{\rm BH}/r_T}\mathcal{R}_{\text{LC}}(E,M_{\rm BH},m_\star) \approx \frac{4Er_T}{GM_{\rm BH}} \ll 1 \label{eq:RLC}
\end{equation}
it could be deduced that the mean $\mathcal{R}$-integrated distribution function has the limiting behaviour of $\mathcal{R}\approx1$: 
\begin{equation}
    \Bar{f}(E) = \frac{\int^{1}_{\mathcal{R}_{\text{LC}}}f(E,\mathcal{R})\rm{d}\mathcal{R}}{\int^{1}_{\mathcal{R}_{\text{LC}}}\rm{d}\mathcal{R}} \approx f(E,1) \label{eq:fE}
\end{equation}
The limit $E\ll GM_{\rm BH}/r_T$ is used here since the majority of stars approach in elliptical orbits with semi-major axes $r_a\gg r_T$.  

The rate calculation in this work follows closely that in \citet{2021arXiv210305883P}, which have shown in detail the derivations of TDE rate by assuming: 
\begin{itemize}
    \item A stellar population described by the Kroupa \textbf{stellar mass function} the MS stars \citep{2001MNRAS.322..231K}: 
\begin{equation}
    \frac{\phi(m_\star)}{M_\odot^{-1}} = \phi_0
    \begin{cases}
        \left(\displaystyle\frac{m_\star}{0.5\,M_\odot}\right)^{-1.3}&\text{for}\,\, 0.08\,M_\odot \leq m_\star \leq 0.5\,M_\odot \\
        \left(\displaystyle\frac{m_\star}{0.5\,M_\odot}\right)^{-2.3}&\text{for}\,\, 10\,M_\odot \leq m_\star \leq 0.5\,M_\odot \\
        0 &\text{else}
    \end{cases} \label{eq:KroupaMF}
\end{equation}
The normalization constant $\phi_0$ ensures that $\displaystyle\int^{\infty}_{0}\phi(m_\star)\rm{d}m_\star=1$. 

\item A power-law \textbf{stellar density profile} near the SMBH: 
\begin{equation}
    \rho(r) = \rho_0\left(\frac{r}{r_{\text{inf}}}\right)^{-\alpha} \label{eq:stellardensity}
\end{equation}
where $r_{\rm inf}$ is the SMBH radius of influence, defined as the radius where $2\pi\displaystyle\int^{r_{\rm inf}}_{0}u^2\rho(u)\rm{d}u=M_{\rm BH}$.  We assume an isothermal profile $(\alpha=2)$, which leads to 
\begin{equation}
    r_{\text{inf}} = \frac{GM_{\rm BH}}{2\sigma^2} \label{eq:rinf}
\end{equation}
where $\sigma$, the velocity dispersion of the galaxy, is constant in the case of an isothermal sphere.  
\item An \textbf{$M_{\rm BH}-\sigma$ black hole mass scaling relation} by \citet{2001ASPC..249..335M}: 
\begin{equation}
    \sigma = 200\left(\frac{M_{\rm BH}}{(1.48\pm0.24)\times10^8\,M_\odot}\right)^{1/(4.65\pm0.48)}\,\text{km s}^{-1} \label{eq:Kor_Msigma}
\end{equation}
\end{itemize}

Based on the above assumptions, the \textbf{differential TDE rate} as a function of $(M_{\rm BH},\beta,m_\star)$ after careful consideration of stellar diffusion through two-body gravitational scattering is given by \citet{2021arXiv210305883P} and \citet{2011PhDT.......385S}: 
\begin{align}
    \mathcal{G}(E,M_{\rm BH},\beta,m_\star)\,\, =\,\, &\frac{f(E)}{1+q^{-1}\xi\ln\left(\mathcal{R}^{-1}_{\text{LC}}\right)}\times \label{eq:diffrate2} \\
    &\left[1 - 2\sum^{\infty}_{m=1}\frac{e^{-\alpha_m^2q/4}}{\alpha_m}\frac{J_0\left(\alpha_m\beta^{-1/2}\right)}{J_1(\alpha_m)}\right], \nonumber \\
    \xi\,\, =\,\, &1 - 4\sum^\infty_{m=1}\frac{e^{-\alpha_m^2q/4}}{\alpha_m^2}, \label{eq:diffrate3} \\
    q(E,M_{\rm BH},m_\star)\,\, =\,\, &\frac{P(E)\Bar{\mu}(E,\langle m_\star^2\rangle^{1/2})}{\mathcal{R}_{\text{LC}}}, \label{eq:diffrate4}
\end{align}
where $q$ and $\bar{\mu}$ are the LC filling factor and orbit-averaged diffusion coefficient respectively.  $J_0$ and $J_1$ are the Bessel functions of the first kind of order zero and one respectively, and $\alpha_m$ is the $m$-th zero of $J_0$.  

By assuming a power-law stellar density profile around the SMBH, the stellar distribution function $f$ and the LC re-filling factor $q$ could be computed analytically rather than numerically in the usually complicated scenarios.  The expressions are as follows \citep{1999MNRAS.309..447M, 2011PhDT.......385S, 2013CQGra..30x4005M, 2016MNRAS.455..859S}: 
\begin{align}
    f(E) &= \left(2\pi\sigma_{\text{inf}}^2\right)^{-3/2}\frac{\rho_0}{\langle m_\star\rangle}\frac{\gamma(\alpha+1)}{\gamma(\alpha-1/2)}\left(\frac{E}{\sigma_{\text{inf}}^2}\right)^{\alpha-3/2}, \\
    q(E,m_\star) &= \nu\left(\frac{E}{\sigma_{\text{inf}}^2}\right)^{\alpha-4}, \\
    \nu(m_\star) &= \frac{8\sqrt{\pi}}{3}(3-\alpha)\frac{\gamma(\alpha+1)}{\gamma(\alpha-1/2)}\left[\frac{5}{32(\alpha-1/2)} + \frac{3I_B(1/2,\alpha) - I_B(3/2,\alpha)}{4\pi}\right]\left(\frac{G\langle m_\star^2\rangle}{\sigma_{\text{inf}}^2r_T}\right)\ln\Lambda, 
\end{align}
where $\sigma_{\text{inf}} = \sqrt{GM_{\rm BH}/r_{\text{inf}}}$ is the velocity dispersion at the radius of influence of BH with mass $M_{\rm BH}$, $\ln\Lambda = \ln(0.4M_{\rm BH}/\langle m_\star\rangle)$ is the Coulomb logarithm, $\gamma$ is the Euler-Gamma function, and $I_B$ is an expression defined in \citet{2011PhDT.......385S}: 
\begin{equation}
    I_B\left(\frac{n}{2},\alpha\right) = \int^1_0 t^{-\frac{n+1}{2}}(1-t)^{3-\alpha}B\left(t,\frac{n}{2},\alpha-\frac{1}{2}\right)\rm{d}t, 
\end{equation}
where $B\left(t,\displaystyle\frac{n}{2},\alpha-\displaystyle\frac{1}{2}\right)$ is the incomplete Euler-Beta function.  

Initially, the orbit-averaged diffusion coefficient $\bar{\mu}$ in Eq.\ref{eq:diffrate4} should be a function of the specific energy of the disrupted star $E$, the mass of the star $m_\star$ and the mass of a background star $m_{\text{bg}}$, assuming a monochromatic distribution of stellar mass.  \citet{1999MNRAS.309..447M} show that the derived rate using a stellar population is the same as a monochromatic distribution of stellar masses of $m_{\text{bg}}=\langle m_\star^2\rangle^{1/2}$, where $\langle m_\star^2\rangle=\int m_\star^2\phi(m_\star)\rm{d}m_\star$.  \citet{2021arXiv210305883P} ignores the effect of mass segregation in which this approximation is no longer valid when stellar distribution becomes a function of position.  

In order to apply a semi-analytical calculation of TDE rate based on loss cone dynamics onto a general galaxy system, numerous assumptions and approximations have to be made.  For a more accurate estimate of the TDE rate in a specific galaxy, characteristics of the galaxy, such as the stellar density profile and the host black hole mass, shall not simply be assumed to follow Eq.\ref{eq:KroupaMF}, \ref{eq:stellardensity}, and \ref{eq:Kor_Msigma}.

\section{Effect from Uncertainty in Black Hole Mass Function} \label{appsec:BHMF}
\noindent One might suspect that the suppression of the TDE host $M_{\rm BH}$ demographics at low-mass end could result from the large uncertainty of BHMF at $M_{\rm BH} \lesssim 10^6\,M_\odot$. For example, the $2\sigma$ uncertainty regions in Fig.2 of \citet{2019ApJ...883L..18G} indicate that the volumetric density of SMBHs might increase or decrease with decreasing $M_{\rm BH}$ in this mass range.  In the lower-mass end ($M_{\rm BH}\sim10^{4-5}\,M_\odot$), the maximum spread is about $^{+0.5\rm{dex}}_{-0.8\rm{dex}}$.

\begin{figure*}
\centering
\includegraphics[width=1.0\linewidth]{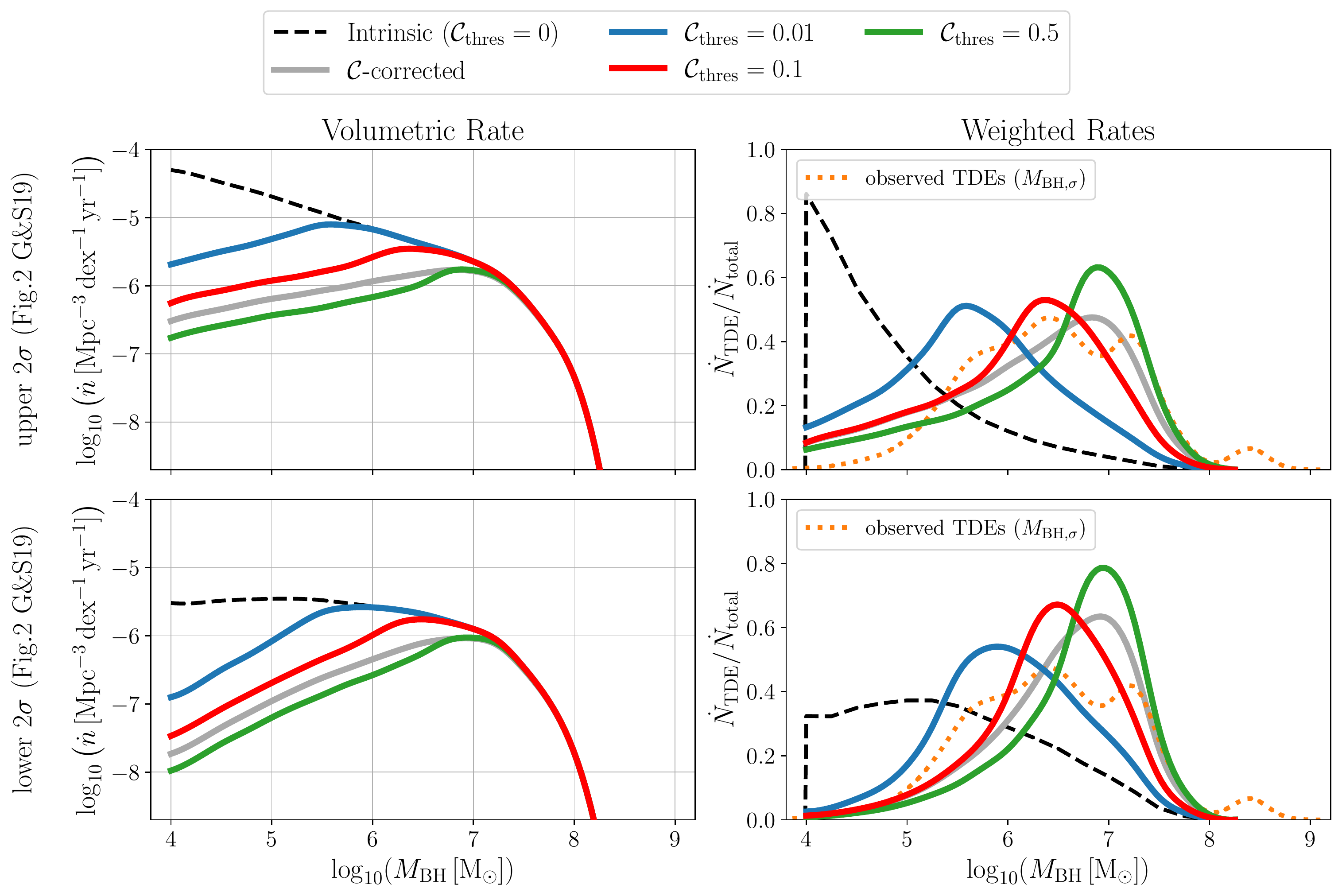}
\caption{\textbf{Regenerated Fig.\ref{fig:totalrate} using the 95\% confidence regions of BHMF Fig.2 in \citet{2019ApJ...883L..18G}.}  The upper (lower) panel corresponds to using the BHMF upper (lower) margin of the light yellow areas of \citet{2019ApJ...883L..18G} Fig.2.  The distributions with $\mathcal{C}_{\text{thres}}=0.1$ (red solid) still best replicate the median of the observed TDE samples.  The rate suppression around light SMBHs due to slow disk formation is more dominant than that by the lower $2\sigma$ of the unconstrained BHMF. \label{fig:BHMF_totalrate}}
\end{figure*}

We calculate the corresponding TDE volumetric rates and their host $M_{\rm BH}$ distributions using the upper and lower 95\% confidence regions in \citet{2019ApJ...883L..18G} (Fig.\ref{fig:BHMF_totalrate}).  The overall shapes of the intrinsic and the $\mathcal{C}$-corrected distributions still remain somewhat similar to that of Fig.\ref{fig:totalrate}, despite some boost (upper $2\sigma$) or further suppression (lower $2\sigma$) at the low-mass ends.  It can be seen that even with the lower $2\sigma$ BHMF, the intrinsic TDE black hole mass distribution still has too much weight towards the lower side.  In order to shift the theoretical distributions to peak at that of the observed, $\mathcal{C}_{\rm thres}\approx0.1$ is still needed in both cases.

\end{document}